\documentclass[journal]{IEEEtran}
\usepackage{cite}
\ifCLASSINFOpdf
\else
\fi
\usepackage[ruled,linesnumberedhidden]{algorithm2e}

\usepackage{color}
\usepackage{amsmath,amssymb,amsfonts}
\usepackage[pdftex]{graphicx}
\makeatletter
\newcommand{\rmnum}[1]{\romannumeral #1}
\newcommand{\Rmnum}[1]{\expandafter\@slowromancap\romannumeral #1@}
\makeatother
\usepackage{algorithmicx} 
\usepackage{algpseudocode}  
\usepackage{amsmath,amssymb,amsfonts}
\usepackage{array}
\usepackage{stfloats}
\begin{document}
\title{Secure and Energy Efficient Transmission for IRS-Assisted Cognitive Radio Networks}
\author{Xuewen Wu,~\IEEEmembership{Graduate Student Member,~IEEE}, Jingxiao Ma, Zhe Xing,~\IEEEmembership{Graduate Student Member,~IEEE}, Chenwei Gu, Xiaoping Xue, 
	 and~Xin Zeng,~\IEEEmembership{Member,~IEEE}
	\thanks{This work was supported by National Key Research and Development Project under Grant 2018YFB0105101, National Natural Science Foundation of China under Grant 61871290, Shanghai Sailing Program under Grant 19YF1451500. (\textit{Corresponding author: Jingxiao Ma}.)}
	\thanks{{X. Wu, J. Ma, Z. Xing, X. Xue, and X. Zeng are with the College of Electronic and Information Engineering, Tongji University, Shanghai 201804, China. (email: wuxuewen1995@163.com;  mjxiao@tongji.edu.cn; zxing@tongji.edu.cn; xuexp@tongji.edu.cn; zengxin1@tongji.edu.cn).}}
	\thanks{{C. Gu is with the College of Telecommunications and Information Engineering, Nanjing University of Posts and Telecommunications, Nanjing 210003, China. (email: 15706290291@139.com).}}}

\markboth{Journal of \LaTeX\ Class Files,~Vol.~14, No.~8, March~2021}%
{Shell \MakeLowercase{\textit{et al.}}: Bare Demo of IEEEtran.cls for IEEE Journals}

\maketitle

\begin{abstract}
	The spectrum efficiency (SE) and security of the secondary users (SUs) in the cognitive radio networks (CRNs) have become two main issues due to the limitation interference to the primary users (PUs) and the shared spectrum with the PUs. Intelligent reflecting surface (IRS) has been recently proposed as a revolutionary technique which can help to enhance the SE and physical layer security of wireless communications. This paper investigates the application of IRS in an underlay CRN, where a multi-antenna cognitive base station (CBS) utilizes spectrum assigned to the PU to communicate with a SU via IRS in the presence of multiple coordinated eavesdroppers (Eves). To achieve the trade-off between the secrecy rate (SR) and energy consumption, we investigate the secrecy energy efficiency (SEE) maximization problem by jointly designing the transmit beamforming at the CBS and the reflect beamforming at the IRS. To solve the non-convex problem with coupled variables, we propose an iterative alternating optimization algorithm to solve the sub-problems alternately, by utilizing an iterative penalty function based algorithm for sub-problem 1 and the difference of two-convex functions method for sub-problem 2. Furthermore, we provide a second-order-cone-programming (SOCP) approximation approach to reduce the computational complexity. Finally, the simulation results demonstrate that IRS can help significantly improve the SE and enhance the physical layer security in the CRNs. Moreover, the effectiveness and superiority of our proposed algorithm in achieving the trade-off between the SR and energy consumption are verified.
\end{abstract}


\IEEEpeerreviewmaketitle

\section{Introduction}
\IEEEPARstart{C}{ognitive} radio network (CRN) has been proposed as an effective way to enhance the spectrum efficiency (SE) \cite{5}. In a typical underlay CRN, secondary users (SUs) can use the spectrum authorized to the primary users (PUs), unless the PU's quality of service (QoS) is crucially affected \cite{re9}. However, there are two main problems in CRNs. One problem is that the SE of the SUs is limited by the interference temperature (IT) constraint imposed on PUs, meaning that the performance improvements for the PU and the SU are conflicting \cite{a13}--\cite{a16}. Specifically, increasing the transmit power at the cognitive base station (CBS) to enhance the signal strength will bring increased interference towards the PU. In this situation, many power allocation and beamforming approaches have been studied to support the optimal transmission \cite{a21}--\cite{a23}. Besides, in order to satisfy the interference limitation constraint, it is another approach to make the SU signals aloof from the PU by adopting multiple antennas. However, the performance improvement of these approaches is limited when the direct link from the CBS to the SU is blocked or very weak. In addition, spectrum management is an effective way from another perspective to enhance SE in CRNs. Based on this, the authors of \cite{TT1} and \cite{TT2} investigated the learning-based spectrum management on cognitive radio network. However, neither \cite{TT1} nor \cite{TT2} considers the security issue, and it is relatively difficult to implement spectrum management to deal with security issues in CRNs.

The other problem is that the characteristics of cognitive radios can introduce security threats and challenges in networks. Compared with the traditional wireless networks without using cognitive radios, security issue in the CRNs becomes more complex since SU is allowed to share the spectrum with the PU \cite{a17}--\cite{a20}. Users within the coverage area of the SU’s transmitter can eavesdrop the confidential information. To deal with the problem, some physical layer security (PLS) technologies can be utilized to ensure the secure transmission. The key point of PLS lies in that when the transmission rate of the legitimate link is greater than that of the wiretap link, good security can be achievable and the rate difference between the legitimate link and the wiretap link is defined as the secrecy rate (SR) \cite{add}. In order to further promote the secure transmission, some technologies have been put forward and combined with PLS to reduce the security risks of eavesdropping, such as cooperative relaying \cite{a27}, beamforming \cite{a24}, zero-forcing-based beamforming \cite{a25}, and artificial noise (AN) injection \cite{a26}. However, these existing approaches exist two main drawbacks. First, deploying active relays or other auxiliary helpers for security transmission will lead to high hardware cost and consumes additional energy. Second, in the adverse wireless transmission environment, it is difficult to ensure satisfactory secrecy performance even if AN or jamming signals are used.

The aforementioned problems in the CRNs mentioned above can be well solved by introducing an intelligent reflecting surface (IRS) into CRNs. IRS has received significant attention from both academia and industry as a promising technology to significantly increase the energy efficiency (EE) and SE in 6G communications \cite{1}, due to full-duplex transmission and low power consumption. It is a new cost-effective and energy efficient technology which is capable of shaping the radio propagation environment and is very suitable for the case in which the direct link from the CBS to the SU is blocked or very weak. The signals reflected by IRS can be superimposed with the signals from the line-of-sight (LoS) paths to enhance the desired signal power at the SUs, and destructively with the signals from the LoS paths to reduce the signal power received at the PUs and eavesdroppers (Eves) through jointly optimizing the transmit beamforming at the CBS and the reflect beamforming at the IRS. The existing contributions have demonstrated benefits brought by introducing an IRS into wireless communication systems. For example, some certain performance criterions such as channel capacity or received signal power \cite{a1}--\cite{3}, secrecy rate \cite{a6}--\cite{re7}, transmit power \cite{a10}--\cite{4}, energy efficiency \cite{a12} are effectively optimized by jointly designing the transmit beamforming at the base station (BS) and the reflect beamforming at the IRS.

However, there is a paucity of research in the transmit and reflect beamforming design for the IRS-assisted CRNs so far. In \cite{a28}, the authors investigated the downlink transmit power minimization problem for the IRS-assisted single-cell CRN coexisting with a single-cell primary radio network, by jointly optimizing the transmit beamformers at the SU transmitter and the phase shift matrix at the IRS. The authors of \cite{a29} investigated the robust beamforming design based on the statistical channel state information (CSI) error model for PU-related cascaded channels in IRS-assisted CRNs to minimize the SU's total transmit power. Since the power minimization problem may be infeasible due to the conflicting constraints of SU’s QoS requirements and PU’s limited interference imposed by the SU, the achievable rate of SU maximization problem subject to the total transmit power constraint of CBS and the interference temperature constraints of PUs has been studied in \cite{a30}--\cite{a32}.

It should be pointed out that the aforementioned works in the IRS-assisted CRNs aimed at maximizing the transmission rate of SUs \cite{a30}--\cite{a32} or minimizing the transmit power \cite{a28}--\cite{a29}. All of these related works in the IRS-assisted CRNs ignored the security issue. Based on this, the authors of \cite{re8} studied an IRS-assisted spectrum sharing underlay cognitive radio wiretap channel, and aimed at enhancing the secrecy rate of SU in this channel. However, greedily pursuing the optimization of transmission rate \cite{a30}--\cite{re8} could likely lead to excessive energy consumption, which is detrimental to limited energy devices. Similarly, greedily pursuing the minimization of transmit power \cite{a28}--\cite{a29} may in turn affect the transmission rate. Thus, both the security issue and the mutual restriction issue of transmission rate and power consumption should be considered. It is imperative to balance the secrecy rate and the energy consumption. To this end, secrecy energy efficiency (SEE), defined as the ratio of the SR to the total power consumption, has been proposed in \cite{v1} to evaluate the available secret bits per unit energy cost. To our best knowledge, we have not found related studies yet in SEE design in the IRS-assisted CRNs. This observation motivates our work in this paper. In summary, our main contributions are listed as follows:

\begin{itemize}
	\item This is the first research on studying the performance trade-off between the energy consumption and secure transmission rate in the IRS-assisted CRNs. Specifically, we propose a new framework to maximize the SEE by jointly optimizing the transmit beamforming at the CBS and the reflect beamforming at the IRS subject to the maximum transmit power of CBS, the minimum SR of SU, the limited interference temperature of PU and the unit modulus constraint of IRS. The problem is challenging to solve due to its non-convexity and coupling of the transmit beamforming at the CBS with the reflect beamforming at the IRS, for which an iterative alternating optimization algorithm is proposed to solve the non-convex problem.
	
	\item In order to optimize the reflect beamforming at the IRS, we introduce an auxiliary variable and convert the original non-convex problem into a semi-definite programming (SDP) problem with rank-1 constraint, and then propose an iterative penalty function based algorithm to implement the optimal reflect beamforming.
	
	\item In order to optimize the transmit beamforming at the CBS, we first convert the original problem into an equivalent subtractive form. Then, as for the rank-1 constraint, we prove that the rank-1 optimal solution always exists. After relaxing the rank-1 constraint, we transform the equivalent subtractive form objective function into a convex optimization function by employing the difference of convex functions (D.C.) method. Furthermore, to reduce the computational complexity, we provide a second-order-cone-programming (SOCP) approximation approach by introducing an auxiliary variable to transform the logarithmic function into the linear function.
	
	\item 	The simulation results show that IRS can help significantly improve the SE and enhance the physical layer security in the CRNs and demonstrate that the proposed algorithm can achieve the highest SEE among all the benchmark methods, indicating that our algorithm can achieve a good trade-off between the SR and energy consumption in the IRS-assisted CRNs. Moreover, we address that there exists a critical value for the minimum acceptable SR threshold of our proposed algorithm, which further indicates that both the SEE and SR of the proposed algorithm can be maximized under the condition that the minimum acceptable SR constraint is met.

\end{itemize} 

The rest of the paper is organized as follows. Section \uppercase\expandafter{\romannumeral2} introduces the system model followed by the problem formulation. The optimization problem is decoupled into two sub-problems, and is solved in section \uppercase\expandafter{\romannumeral3}. Section \uppercase\expandafter{\romannumeral4} provides a SOCP approach to reduce the computational complexity of the algorithm that solve the transmit beamforming at the CBS. The simulation results are presented in Section \uppercase\expandafter{\romannumeral5}, and this paper is concluded in Section \uppercase\expandafter{\romannumeral6}.

\textit{Notations}: Vectors and matrices are represented by lowercase and uppercase bold typeface letters, respectively. ${\left(\cdot \right)^T}$ and ${\left(\cdot \right)^H}$ indicate the transpose and Hermitian transpose operation, respectively. $\left\| {\cdot} \right\|$ represents the Euclidean norm of a vector. $\left|\cdot\right|$ is the absolute value. $tr({\bf{X}})$, $rank({\bf{X}})$, $\left| {\bf{X}} \right|$ and $\lambda_{\max}({\bf{X}})$ denote the trace, the rank, the determinant and the maximum eigenvalue of matrix $\bf{X}$, respectively. $diag({\bf{x}})$ represents the diagonal matrix with ${\bf{x}}$ on its main diagonal. ${\bf{I}}_N$ is the $N\times N$ identity matrix. ${\left[ {\bf{X}} \right]_{i,j}}$ is the $\left( {i,j} \right)$-th element of $\bf{X}$. ${\bf{X}}{\underline \succ} 0$ indicates that $\bf{X}$ is a positive semi-definite matrix. $\left\langle {{\bf{X}},{\bf{Y}}} \right\rangle  = tr\left( {{{\bf{X}}^H}{\bf{Y}}} \right)$. ${\log _2}\left( \cdot \right)$ denotes the logarithmic function. $E\left( {\cdot} \right)$ represents the expectation operator. ${\cal C}^{M \times N}$ stands for the complex space of $M\times N$.

\section{System Model and Problem Formulation}

As shown in Fig.1, we consider an IRS-assisted CRN: a CBS utilizes spectrum assigned to the PU to communicate with a SU via IRS, where \textit{K} coordinated Eves attempt to intercept the CBS-SU transmission. Suppose that CBS is equipped with \textit{N} antennas, and SU, PU, and Eves own single antenna. For enhancing the PLS and SE of the CRN, an IRS is deployed on the facade of a tall building. The IRS is composed of \textit{L} passive reflecting elements, each of which can flexibly adjust the phase of the incident electromagnetic wave. Due to the large path loss, the power of the signals reflected by IRS twice or more is negligible \cite{a31}.

All channels in our considered network are supposed to undergo quasi-static flat-fading. The channel coefficients from the CBS to IRS, from the CBS to SU, from the CBS to PU, from the CBS to the \textit{k}-th Eve, from the IRS to SU, from the IRS to PU, and from the IRS to the \textit{k}-th Eve are denoted as ${{\bf{H}}_{CI}} \in {{\cal{C}}^{L \times N}}$,  ${{\bf{h}}_{CS}} \in {{\cal{C}}^{N \times 1}}$, ${{\bf{h}}_{CP}} \in {{\cal{C}}^{N \times 1}}$, ${{\bf{h}}_{CEk}} \in {{\cal{C}}^{N \times 1}}$,  ${{\bf{h}}_{IS}} \in {{\cal{C}}^{L \times 1}}$, ${{\bf{h}}_{IP}} \in {{\cal{C}}^{L \times 1}}$ and ${{\bf{h}}_{IEk}} \in {{\cal{C}}^{L \times 1}}$. In addition, by assuming the scenario that the Eves are active users yet untrusted by the legitimate user, the CSIs of the wiretap links can be obtained \cite{9}. The computation of resource allocation is executed in CBS, and then the CBS needs to convey the resource allocation results (reflection matrix of the IRS) to IRS, whose phased shifts are controlled by an attached controller. Therefore, the CBS can transmit the reflection matrix to controller via a dedicated separate wireless control link. In order to describe the performance limit of the IRS-assisted CRN, we suppose that perfect CSIs {\footnote{ Usually, there are two main methods for the IRS-involved channel acquisition, relying on whether the IRS elements are equipped with receive RF chains or not \cite{1}. For the first method with receive RF chains, conventional channel estimation methods can be applied for the IRS to estimate the channels of the CBS-IRS and IRS-user links, respectively. For the second method without receive RF chains at the IRS, the IRS reflection patterns can be designed together with the uplink pilots to estimate the cascaded CBS-IRS-user channels \cite{re4}-\cite{re5}.} }of all channels are available.
\begin{figure}[t]
	\centering
	\includegraphics[width=8cm]{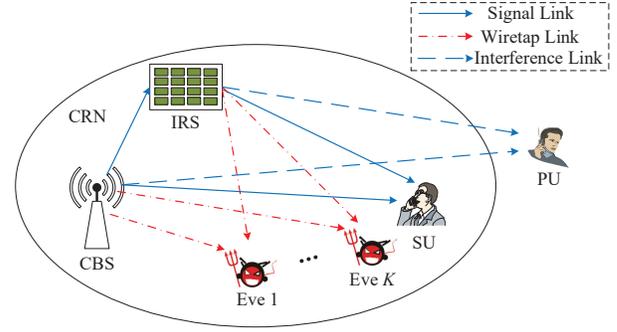}
	\caption{An IRS-assisted cognitive radio network.}
	\label{fig1}
\end{figure}

When the CBS transmits signal $x$ to SU via IRS. The interference from primary base station (PBS) to SU and Eves can be regarded as noises due to the long distance between PBS and CRN. The signals received at the SU, PU, and the \textit{k}-th Eve can be uniformly written as
\begin{equation}
	{y_v} = \left( {{\bf{h}}_{Iv}^H{\bf{Q}}{{\bf{H}}_{CI}} + {\bf{h}}_{Cv}^H} \right){\bf{w}}x + {n_v},\space v \in \left\{ {S,P,Ek} \right\},
\end{equation}where $x$ is the transmit signal following $E\left( {{{\left| x \right|}^2}} \right) = 1$. $\bf{w}$ represents the transmit beamforming at the CBS. ${\bf{Q}}{\rm{ }} = diag\left( {{\beta _1}{e^{j{\theta _1}}}, \cdots ,{\beta _l}{e^{j{\theta _l}}}, \cdots ,{\beta _L}{e^{j{\theta _L}}}} \right)$ is the phase shift matrix of the IRS. $\theta _l$ and ${\beta _l} \in \left[ {0,1} \right]$ are the phase shift and the amplitude reflection coefficient of the \textit{l}-th reflecting element, respectively. Theoretically, the reflection amplitude of each element can be adjusted for different purposes such as channel estimation, energy harvesting, and performance optimization \cite{1}. However, in practice, it is costly to implement independent control of the reflection amplitude and phase shift simultaneously. Thus, each element is usually designed to maximize the signal reflection for simplicity \cite{a3}, \cite{re1}-\cite{re3}. As such, we assume ${\beta _l} = 1,\forall l$, in the sequel of this paper. $n_v$ is the additive complex white Guassian noise, in which the entries are with zero-mean and variance $\sigma _v^2$. 

Briefly, we denote ${{\bf{H}}_v} = \left[ {\begin{array}{*{20}{c}}
		{diag\left( {{\bf{h}}_{Iv}^H} \right){{\bf{H}}_{CI}}}\\
		{{\bf{h}}_{Cv}^H}
\end{array}} \right]$, $v \in \left\{ {S,P,Ek} \right\}$. Therefore, (1) can be further expressed as 
\begin{equation}
	{y_v} = {\bf{q}}_{}^H{{\bf{H}}_v}{\bf{w}}x + {n_v},
\end{equation}where ${\bf{q}} \buildrel \Delta \over = {\left[ {{e^{j{\theta _1}}},{e^{j{\theta _2}}}, \cdots ,{e^{j{\theta _L}}},1} \right]^H}$. Accordingly, the signal-to-noise ratio (SNR) at the receiver can be expressed as
\begin{equation}
	{\gamma _v} = \frac{{{{\left| {{\bf{q}}_{}^H{{\bf{H}}_v}{\bf{w}}} \right|}^2}}}{{\sigma _v^2}}.
\end{equation}

Assuming that \textit{K} Eves eavesdrop on the signal sent by the CBS coordinately. The achievable SR at the SU can be expressed as
\begin{equation}
	\begin{split}
		{R_{sec }} = { {{R_S} - {R_E}}  } = { {{{\log }_2}\left( {1 + {\gamma _S}} \right) - {{\log }_2}\left( {1 + \sum\limits_{k = 1}^K {{\gamma _{Ek}}} } \right)}  }\\
		={\log _2}\left( {1 + \frac{{{{\left| {{\bf{q}}_{}^H{{\bf{H}}_S}{\bf{w}}} \right|}^2}}}{{\sigma _S^2}}} \right) - {\log _2}\left( {1 + \sum\limits_{k = 1}^K {\frac{{{{\left| {{\bf{q}}_{}^H{{\bf{H}}_{Ek}}{\bf{w}}} \right|}^2}}}{{\sigma _{Ek}^2}}} } \right).
	\end{split}
\end{equation}
where $R_S$ and $R_E$ represent the transmission rate at SU and Eve, respectively.

The energy consumed by the CBS includes the transmit power $\left\| {\bf{w}} \right\|^2$ and the circuit power $P_{CBS}$. Denote the power consumed by the IRS as ${P_{IRS}} = {P_{Sta}} + L{P_{Dyn}}$ \cite{add10}, where ${P_{Sta}}$ and ${P_{Dyn}}$ are the static power required to maintain the basic circuit operations of the IRS and the dynamic power per reflecting component, respectively. Generally, ${P_{Dyn}}$ is much smaller than ${P_{Sta}}$. Since we focus on small IRS, $L{P_{Dyn}}$ can be ignored and thereby $P_{IRS}$ can be set a constant, i.e., ${P_{IRS}} = {P_{Sta}}$. As such, the total power consumption of the considered network is expressed as \cite{a12}
\begin{equation}
	{P_{tot}} = \zeta \left\| {\bf{w}} \right\|^2 + {P_{CBS}} + {P_{IRS}},
\end{equation}where $\zeta$ represents the amplifier coefficient.

In order to keep a balance between the SR and total power consumed by the system, we employ SEE \cite{v1} as the performance metric to calculate the secret bits per unit energy and bandwidth during the transmission, which can be expressed as
\begin{equation}
	\eta_{SEE}  = \frac{{{R_{sec }}}}{{{P_{tot}}}}({\text{bit/Joule/Hz}}).
\end{equation}

At the same time, during the cognitive transmission, the interference temperature to PU must be lower than a predefined threshold $I_p^{th}$ to ensure the QoS of PU, i.e.,
\begin{equation}
	{I_p} = {\left| {{\bf{q}}_{}^H{{\bf{H}}_P}{\bf{w}}} \right|^2}\le I_p^{th}.
\end{equation}

Based on the above description, we establish a SEE maximization (SEE-Max) problem under the premise that the secure transmission requirement of SU and the transmit power of cognitive transmission are ensured, the leakage interference to PU restricted under a predefined threshold, as well as the unit modulus constraint of IRS, namely,
\begin{align}
	\max\limits_{{\bf{w}},{\bf{q}}}\quad &\frac{{{R_{sec}}}}{{\zeta \left\| {\bf{w}} \right\|^2 + {P_{CBS}} + {P_{IRS}}}} \tag{8a}\\
	{\text{ s.t.}}\quad&R_{sec}\ge R_{sec }^{min }, \tag{8b}\\
	&\left| {{{\left[ {\bf{q}} \right]}_l}} \right| = 1,\forall l \in \left\{ {1, \cdots ,L + 1} \right\}=\mathbb L, \tag{8c}\\
	&{\left| {{\bf{q}}_{}^H{{\bf{H}}_P}{\bf{w}}} \right|^2} \le I_p^{th} \quad {\text{and}} \quad \left\| {\bf{w}} \right\|^2 \le P_c^{\max }, \tag{8d}
\end{align}where ${{{\left[ {\bf{q}} \right]}_l}}$ denotes the \textit{l}-th element of $\bf{q}$. $\mathbb{L}$ represents the set for all $l$. $R_{sec}^{min}\ge0$ stands for the minimum acceptable SR threshold, aiming to ensure certain level of SU's secure transmission according to the user's requirement. From the perspective of information theory, if the channel instantaneous SR is larger than the minimum acceptable SR threshold, the required secrecy level will be guaranteed. Otherwise, the confidentiality of information secrecy will be threaten. $P_c^{max}$ represents CBS's maximum transmit power.

By means of two new variables ${\bf{\Theta }} = {\bf{q}}{{\bf{q}}^H}$ and ${\bf{W}} = {\bf{w}}{{\bf{w}}^H}$, problem (8) thereby can be equivalently expressed as
\begin{align}
	\max\limits_{{\bf{W}},{\bf{\Theta}}}\quad &\frac{{{R_{\sec }}\left( {{\bf{W}},{\bf{\Theta }}} \right)}}{{\zeta tr\left( {\bf{W}} \right) + {P_{CBS}} + {P_{IRS}}}} \tag{9a}\\
	{\text{ s.t.}}\quad&{{R_{\sec }}\left( {{\bf{W}},{\bf{\Theta }}} \right)} \ge R_{sec }^{min }, \tag{9b}\\
	&{\bf{\Theta }}{\underline \succ}0,\quad rank({\bf{\Theta}})=1,\quad{\rm{  and  }}\quad{\left[ {\bf{\Theta }} \right]_{l, l}} = 1, \forall l \in {\mathbb L}, \tag{9c}\\
	&{\rm{      }}{\bf{W}}{\underline \succ}0,\quad rank({\bf{W}})=1,\quad{\rm{  and  }}\quad tr\left( {\bf{W}} \right) \le P_c^{\max }{\rm{   }}, \tag{9d}\\
	&tr\left( {{\bf{\Theta }}{{\bf{H}}_P}{\bf{WH}}_P^H} \right) \le I_p^{th},\tag{9e}
\end{align} where 
\begin{equation}\nonumber
\begin{array}{l}
{R_{\sec }}\left( {{\bf{W}},{\bf{\Theta }}} \right) = {\log _2}\left( {1 + \frac{{tr\left( {{\bf{\Theta }}{{\bf{H}}_S}{\bf{WH}}_S^H} \right)}}{{\sigma _S^2}}} \right) 
\\\quad\quad\quad\quad\quad\quad\quad- {\log _2}\left( {1 + \sum\limits_{k = 1}^K {\frac{{tr\left( {{\bf{\Theta }}{{\bf{H}}_{Ek}}{\bf{WH}}_{Ek}^H} \right)}}{{\sigma _{Ek}^2}}} } \right).
\end{array}
\end{equation}
 Obviously, both of the two new variables ${\bf{\Theta }}$ and ${\bf{W}}$ are rank-1 symmetric positive semi-definite matrices. Note that problem (9) is non-convex over coupled $\bf{\Theta}$ and $\bf{W}$, making it challenging to deal with problem (9).

\section{Iterative Alternating Optimization Algorithm}
In this section, we propose an iterative alternating optimization algorithm to optimize $\bf{\Theta}$ and $\bf{W}$ alternatively by fixing the other as constant. Thus, problem (9) is decoupled into two sub-problems, i.e., (10) with given $\bf{W}$ and (18) with given $\bf{\Theta}$.

\subsection{Sub-Problem 1: Optimizing $\bf{\Theta}$ With Given $\bf{W}$}
Since $\bf{W}$ has been fixed, we can ignore the denominator part of the objective function (9a). Thus, we can write sub-problem 1 as
\setcounter{equation}{9}
\begin{equation}
	\begin{split}
		\max\limits_{{\bf{\Theta}}}\quad &{\log _2}\left( {1 + \frac{{tr\left( {{\bf{\Theta }}{{\bf{H}}_S}{\bf{WH}}_S^H} \right)}}{{\sigma _S^2}}} \right) \\
		&- {\log _2}\left( {1 + \sum\limits_{k = 1}^K {\frac{{tr\left( {{\bf{\Theta }}{{\bf{H}}_{Ek}}{\bf{WH}}_{Ek}^H} \right)}}{{\sigma _{Ek}^2}}} } \right)\\
		{\text{ s.t.}}\quad &(9b),(9c),(9e).
	\end{split}
\end{equation}Note that (9b) is a constraint on the objective optimization function, and constraint (9b) can be temporarily ignored. We only need to check constraint (9b) after solving sub-problem 1. As such, sub-problem 1 can be further written as
\begin{equation}
	\begin{split}
		\min\limits_{{\bf{\Theta}}}\quad &\frac{{1 + \sum\limits_{k = 1}^K {\frac{{tr\left( {{\bf{\Theta }}{{\bf{H}}_{Ek}}{\bf{WH}}_{Ek}^H} \right)}}{{\sigma _{Ek}^2}}} }}{{1 + \frac{{tr\left( {{\bf{\Theta }}{{\bf{H}}_S}{\bf{WH}}_S^H} \right)}}{{\sigma _S^2}}}}\\
		{\text{ s.t.}}\quad &(9c),(9e).
	\end{split}
\end{equation}Then, by means of two introduced variables $t$ and $a$, a positive semi-definite matrix $\bf{A}$ meeting ${\bf{A}}=a{\bf{\Theta}}$, and employing the Charnes-Cooper transformation \cite{10}, we are able to easily transform the sub-problem 1 into a SDP problem, given by
\begin{align}
	{\mathop {\min }\limits_{{\bf{A}}{\underline \succ}0, a\ge0} \quad }&{t }\tag{12a}\\
	{{\rm{ s}}{\rm{.t}}{\rm{.}}\quad }&{a + \frac{{tr\left( {{\bf{A}}{{\bf{H}}_S}{\bf{WH}}_S^H} \right)}}{{\sigma _S^2}} \ge 1},\tag{12b}\\
	&a + \sum\limits_{k = 1}^K {\frac{{tr\left( {{\bf{A}}{{\bf{H}}_{Ek}}{\bf{WH}}_{Ek}^H} \right)}}{{\sigma _{Ek}^2}}}  \le t,\tag{12c}\\
	&tr\left( {{\bf{A }}{{\bf{H}}_P}{\bf{WH}}_P^H} \right) \le aI_p^{th} ,\tag{12d}\\
	&{{{\left[ {\bf{A}} \right]}_{l,l}} = a,\forall l \in {\mathbb L} },\tag{12e}\\
	&rank({\bf{A}}) = 1.\tag{12f}
\end{align}

\textit{Proposition 1}: The optimization problem (12) is completely equivalent to the problem (11), indicating that the solutions of problems (11) and (12) are the same, given by ${{\bf{\Theta }}^*} = {{{{\bf{A}}^*}} \mathord{\left/{\vphantom {{{{\bf{A}}^*}} {{a^*}}}} \right.\kern-\nulldelimiterspace} {{a^*}}}$.

\quad\textit{Proof}: Appendix A.

It is obvious that the constraint (12f) makes the optimization problem hard to solve. In the traditional semi-definite relaxation (SDR) method, (12f) is usually ignored to simplify the problem, and the optimal solution selected among randomly generated rank-1 feasible solutions can be regarded  as an approximately best solution \cite{11}. However, there is probably no optimal solution to the initial SDP problem among the feasible solutions in random space. Even if there is, the chosen rank-1 solution is very likely to be a sub-optimal solution. What's worse, the obtained solution may deviate greatly with the optimal solution. Considering the above issues, we can rewrite (12f) as $rank\left( {\bf{A}} \right) = 1 \Leftrightarrow tr\left( {\bf{A}} \right) - {\lambda _{\max }}\left( {\bf{A}} \right) \le 0$. Then, the constrained problem (12) is further reformulated as
\begin{align}
	{\mathop {\min }\limits_{{\bf{A}}{\underline \succ}0, a\ge0} \quad }&{t}\tag{13a}\\
	{\text{ s.t.}}\quad 
	&(12b)-(12e),\tag{13b}\\
	& tr\left( {\bf{A}} \right) - {\lambda _{\max }}\left( {\bf{A}} \right)\le0.\tag{13c}
\end{align} 

\textit{Proposition 2}: The optimization problem (13) is equivalent to the optimization problem (12).

\quad\textit{Proof}: Appendix B.

Note that the inequality $tr\left( {\bf{A}} \right) \ge {\lambda _{\max }}\left( {\bf{A}} \right) $ always holds for any matrix ${\bf{A}}{\underline \succ}0$. Thus, our purpose is to make $tr\left( {\bf{A}} \right) - {\lambda _{\max }}\left( {\bf{A}} \right)$ as small as possible (approaching zero). With the help of penalty item method, we can incorporate the constraint (13c) into (13a), yielding
\begin{align}
	\min\limits_{{\bf{A}}{\underline \succ}0, a\ge0, t}\quad &{t }+\rho(tr\left( {\bf{A}} \right) - {\lambda _{\max }}\left( {\bf{A}} \right))\tag{14a}\\
	{\text{ s.t.}}\quad
	&(12b)-(12e),\tag{14b}
\end{align}where the penalty coefficient $\rho$ should be large enough to obtain small values of $tr\left( {\bf{A}} \right) - {\lambda _{\max }}\left( {\bf{A}} \right)$. As we can see, the objective function (14a) is concave, making the problem (14) a concave function minimization problem over a convex set, i.e., a concave programming. Moreover, considering that ${\lambda _{\max }}\left( {\bf{A}} \right)$ is a non-smooth function, we can adopt the sub-gradient of the non-smooth function, which is defined as $\partial  {\lambda _{\max }}\left( {\bf{X}} \right) = {{\bf{x}}_{\max }}{\bf{x}}_{\max }^H$. Then, we have \cite{b1}
\setcounter{equation}{14}
\begin{equation}		
	\begin{split}
		{\lambda _{\max }}\left( {\bf{X}} \right) - {\lambda _{\max }}\left( {\bf{A}} \right) \ge \left\langle {{{\bf{a}}_{\max }}{\bf{a}}_{\max }^H,{\bf{X}} - {\bf{A}}} \right\rangle ,\forall {\bf{X}} \ge 0.
	\end{split}
\end{equation}where ${{\bf{a}}_{\max }}$ is the eigenvector corresponding to the maximum eigenvalue of $\bf{A}$.

Then, by employing the maximum eigenvalue as well as the corresponding unit eigenvector ${{\bf{a}}^{\left( n \right)}}$ to initialize the feasible solution ${\bf{A}}^{(n)}$, which satisfies the constraint (14b), a SDP problem can be written as 
\begin{align}
	\min\limits_{{\bf{A}}{\underline \succ}0, a\ge0, t}\quad &{t + \rho \left[ {tr\left( {\bf{A}} \right) - \left\langle {{{\bf{a}}^{(n)}}{\bf{a}}_{}^{(n)H},{\bf{A}}} \right\rangle } \right]}\tag{16a}\\
	{\text{ s.t.}}\quad
	&(12b)-(12e),\tag{16b}
\end{align}
The problem (16) can provide the optimal solution for ${\bf{A}}^{(n+1)}$, which produces a smaller objective value (16a) than that produced by ${\bf{A}}^{(n)}$. In specific, we suppose that ${\bf{A}}^{(n+1)}$ is the optimal solved solution of (16), then we can obtain
\setcounter{equation}{16}
\begin{equation}
	\begin{split}
			F&\left( {{{\bf{A}}^{\left( {n + 1} \right)}}} \right) = t + \rho \left[ {tr\left( {{{\bf{A}}^{\left( {n + 1} \right)}}} \right) - {\lambda _{\max }}\left( {{{\bf{A}}^{\left( {n + 1} \right)}}} \right)} \right]\\
			&\le t + \rho \left[ {tr\left( {{{\bf{A}}^{\left( {n + 1} \right)}}} \right) - {\lambda _{\max }}\left( {{{\bf{A}}^{\left( n \right)}}} \right)} \right.\\
			&\quad\left. { - \left\langle {{{\bf{a}}^{(n)}}{\bf{a}}_{}^{(n)H},{{\bf{A}}^{\left( {n + 1} \right)}} - {{\bf{A}}^{\left( n \right)}}} \right\rangle } \right]\\
			&= t + \rho \left[ {tr\left( {{{\bf{A}}^{\left( {n + 1} \right)}}} \right) - \left\langle {{{\bf{a}}^{(n)}}{\bf{a}}_{}^{(n)H},{{\bf{A}}^{\left( {n + 1} \right)}}} \right\rangle } \right.\\
			&\quad\left. { + \left\langle {{{\bf{a}}^{(n)}}{\bf{a}}_{}^{(n)H},{{\bf{A}}^{\left( n \right)}}} \right\rangle  - {\lambda _{\max }}\left( {{{\bf{A}}^{\left( n \right)}}} \right)} \right]\\
			&\le t + \rho \left[ {tr\left( {{{\bf{A}}^{\left( n \right)}}} \right) - \left\langle {{{\bf{a}}^{(n)}}{\bf{a}}_{}^{(n)H},{{\bf{A}}^{\left( n \right)}}} \right\rangle } \right.\\
			&\quad\left. { + \left\langle {{{\bf{a}}^{(n)}}{\bf{a}}_{}^{(n)H},{{\bf{A}}^{\left( n \right)}}} \right\rangle  - {\lambda _{\max }}\left( {{{\bf{A}}^{\left( n \right)}}} \right)} \right]\\
			&= t + \rho \left[ {tr\left( {{{\bf{A}}^{\left( n \right)}}} \right) - {\lambda _{\max }}\left( {{{\bf{A}}^{\left( n \right)}}} \right)} \right]
			= F\left( {{{\bf{A}}^{\left( n \right)}}} \right),
	\end{split}
\end{equation}
which verifies the iterative procedure.

As a result, we can obtain the optimal solution to problem (16) by means of CVX solvers quickly and accurately. The selection of the penalty coefficient $\rho$ is important for the computational efficiency. Algorithm 1 shows the detailed solution process of sub-problem 1, which includes the choice of the penalty coefficient $\rho$.

\begin{algorithm}[t]
	\caption{The Algorithm for Solving Sub-Problem 1}
	\LinesNumbered 
	\KwIn{${\bf{W}}$, ${\bf{H}}_s$, ${\bf{H}}_p$, ${\bf{H}}_{Ek}$, $P_c^{max}$, $R_{sec}^{min}$, $I_p^{th}$}
	\KwOut{${\bf{\Theta}}^*$}
	{{Initialize}} $n=0$, $t^0=0$, $\rho=10$\; 
	Calculate ${\bf{A}}^{(0)}$ satisfying (14b)\;
	\While{$\left| {tr\left( {{{\bf{A}}^{\left( n \right)}}} \right) - {\lambda _{\max }}\left( {{{\bf{A}}^{\left( n \right)}}} \right)} \right| > \varepsilon $}{
		Find the optimal solution ${\bf{A}}^{(n+1)}$, ${{a}}^{(n+1)}$ and ${{t}}^{(n+1)}$ of problem (16) by using CVX\;
		\eIf{${{\bf{A}}^{\left( {n + 1} \right)}} \approx {{\bf{A}}^{\left( n \right)}}$}{
			Set $\rho:=2\rho$\;
		}{
			Set $n:=n+1$\;
		    Set $\rho=10$\;
		}
	}
	Calculate ${\bf{\Theta}}^* = {\bf{A}}^{(n)}/a^{(n)}$.
\end{algorithm}

\subsection{Sub-Problem 2: Optimizing $\bf{W}$ With Given $\bf{\Theta}$}
Next, we optimize $\bf{W}$ with the solved $\bf{\Theta}$. Sub-problem 2 can be formulated as
\setcounter{equation}{17}
\begin{equation}		
	\begin{split}
		\max\limits_{{\bf{W}}}\quad &\frac{{{R_{\sec }}\left( {{\bf{W}}} \right)}}{{\zeta tr\left( {\bf{W}} \right) + {P_{CBS}} + {P_{IRS}}}}\\
		{\text{ s.t.}}\quad &(9b),(9d),(9e),
	\end{split}
\end{equation}where
\begin{equation}\nonumber
	\begin{array}{l}		 
	{R_{\sec }}\left( {{\bf{W}}} \right) = {\log _2}\left( {1 + \frac{{tr\left( {{\bf{\Theta }}{{\bf{H}}_S}{\bf{WH}}_S^H} \right)}}{{\sigma _S^2}}} \right) 
	\\\quad\quad\quad\quad\quad- {\log _2}\left( {1 + \sum\limits_{k = 1}^K {\frac{{tr\left( {{\bf{\Theta }}{{\bf{H}}_{Ek}}{\bf{WH}}_{Ek}^H} \right)}}{{\sigma _{Ek}^2}}} } \right). 
\end{array}
\end{equation}

By means of an auxiliary variable $\varphi \ge1 $, we can reformulate the sub-problem 2 as
\begin{align}
	\max\limits_{{\bf{W}}{\underline \succ}0,\varphi\ge1}\quad &\frac{{{{\log }_2}\left( {1 + \frac{{tr\left( {{\bf{\Theta }}{{\bf{H}}_S}{\bf{WH}}_S^H} \right)}}{{\sigma _S^2}}} \right) - {{\log }_2}\varphi }}{{\zeta tr\left( {\bf{W}} \right) + {P_{CBS}} + {P_{IRS}}}}\tag{19a}\\
	{\text{ s.t.}}\quad&{\log _2}\left( {1 + \sum\limits_{k = 1}^K {\frac{{tr\left( {{\bf{\Theta }}{{\bf{H}}_{Ek}}{\bf{WH}}_{Ek}^H} \right)}}{{\sigma _{Ek}^2}}} } \right) = {\log _2}\varphi,\tag{19b} \\
	&{\log _2}\left( {1 + \frac{{tr\left( {{\bf{\Theta }}{{\bf{H}}_S}{\bf{WH}}_S^H} \right)}}{{\sigma _S^2}}} \right) - {\log _2}\varphi  \ge R_{sec }^{min },\tag{19c}\\
	&tr\left( {\bf{W}} \right) \le P_c^{\max },\tag{19d}\\
	&tr\left( {{\bf{\Theta }}{{\bf{H}}_P}{\bf{WH}}_P^H} \right) \le I_p^{th},\tag{19e}\\
	&rank\left( {\bf{W}} \right) = 1.\tag{19f}
\end{align}
Rewriting (19b) as ${\log _2}\left( {1 + \sum\limits_{k = 1}^K {\frac{{tr\left( {{\bf{\Theta }}{{\bf{H}}_{Ek}}{\bf{WH}}_{Ek}^H} \right)}}{{\sigma _{Ek}^2}}} } \right) \le {\log _2}\varphi$ does not change the optimal solution of (19), which can be explained as follows: assume that $\left( {{{\bf{W}}^*},{\varphi^*}} \right)$ is the optimal solution satisfying ${\log _2}\left( {1 + \sum\limits_{k = 1}^K {\frac{{tr\left( {{\bf{\Theta }}{{\bf{H}}_{Ek}}{\bf{W}}^*{\bf{H}}_{Ek}^H} \right)}}{{\sigma _{Ek}^2}}} } \right) < {\log _2}\varphi^*$. Then, there definitely exists a certain value $0<\beta<1$, enabling us to chose a feasible point 
 $\left( {{\bf{\bar W}},\bar \varphi} \right) = \left( { {{\bf{W}}^*},{\beta\varphi^*}} \right)$ to make ${\log _2}\left( {1 + \sum\limits_{k = 1}^K {\frac{{tr\left( {{\bf{\Theta }}{{\bf{H}}_{Ek}}{\bf{\bar WH}}_{Ek}^H} \right)}}{{\sigma _{Ek}^2}}} } \right) = {\log _2}\bar \varphi $. Obviously, $\left( {{\bf{\bar W}},\bar \varphi} \right)$ meets the constraints (19c)-(19f) and $\left( {{\bf{\bar W}},\bar \varphi} \right)$ can be proved to provide a larger optimization value (19a) than that provided from $\left( {{{\bf{W}}^*},{\varphi^*}} \right)$, which is in contradiction with the assumption that $\left( {{{\bf{W}}^*},{\varphi^*}} \right)$ is the optimal solution. Therefore, we can rewrite the constraint (19b) as ${\log _2}\left( {1 + \sum\limits_{k = 1}^K {\frac{{tr\left( {{\bf{\Theta }}{{\bf{H}}_{Ek}}{\bf{WH}}_{Ek}^H} \right)}}{{\sigma _{Ek}^2}}} } \right) \le {\log _2}\varphi$, which is a convex constraint after removing the logarithmic sign. Then, (19) can be reformulated as
\begin{align}
	\max\limits_{{\bf{W}}{\underline \succ}0,\varphi\ge1}\quad &\frac{{{{\log }_2}\left( {1 + \frac{{tr\left( {{\bf{\Theta }}{{\bf{H}}_S}{\bf{WH}}_S^H} \right)}}{{\sigma _S^2}}} \right) - {{\log }_2}\varphi }}{{\zeta tr\left( {\bf{W}} \right) + {P_{CBS}} + {P_{IRS}}}}\tag{20a}\\
	{\text{ s.t.}}\quad&\sum\limits_{k = 1}^K {\frac{{tr\left( {{\bf{\Theta }}{{\bf{H}}_{Ek}}{\bf{WH}}_{Ek}^H} \right)}}{{\sigma _{Ek}^2}}}  - \varphi  + 1 \le 0,\tag{20b}\\
	&\frac{{tr\left( {{\bf{\Theta }}{{\bf{H}}_S}{\bf{WH}}_S^H} \right)}}{{\sigma _S^2}} - {2^{R_{sec }^{min }}}\varphi  + 1 \ge 0,\tag{20c}\\
	&tr\left( {\bf{W}} \right) \le P_c^{\max },\tag{20d}\\
	&tr\left( {{\bf{\Theta }}{{\bf{H}}_P}{\bf{WH}}_P^H} \right) \le I_p^{th},\tag{20e}\\
	&rank\left( {\bf{W}} \right) = 1.\tag{20f}
\end{align}

It is not hard to observe that the objective function (20a) is in the form of a fraction, which makes the optimization problem (20) non-convex. To tackle this difficulty, we transform the problem (20) into an equivalent subtractive one by using the Dinkelbach’s method \cite{add1}, given by
\begin{align}
 \max\limits_{{\bf{W}}{\underline \succ}0,\varphi\ge1}\quad &{\log _2}\left( {1 + \frac{{tr\left( {{\bf{\Theta }}{{\bf{H}}_S}{\bf{WH}}_S^H} \right)}}{{\sigma _S^2}}} \right) - {\log _2}\varphi \notag \\
	&- \eta \left( {\zeta tr\left( {\bf{W}} \right) + {P_{CBS}} + {P_{IRS}}} \right)\tag{21a}\\
	{\text{ s.t.}}\quad&(20b)-(20f),\tag{21b}		
\end{align}
where $\eta$ is a non-negative parameter. As for the non-convex rank-1 constraint (20f), we can introduce the proposition 3 to relax it.

\textit{Proposition 3}: Assuming that ${\bf{W}}^*$ is the optimal solution of the optimization problem (21), it always satisfies $rank({\bf{W}}^*)=1$.

\quad\textit{Proof}: Appendix C.

By applying the proposition 3, the rank-1 constraint (20f) can be ignored since the optimal solution of problem (21) always satisfies the rank-1 condition. Thus, by defining 
\setcounter{equation}{21}
\begin{equation}
	\begin{split}
		{f_1}\left( {{\bf{W}},\eta } \right) =& {\log _2}\left( {1 + \frac{{tr\left( {{\bf{\Theta }}{{\bf{H}}_S}{\bf{WH}}_S^H} \right)}}{{\sigma _S^2}}} \right) \\
		&- \eta \left( {\zeta tr\left( {\bf{W}} \right) + {P_{CBS}} + {P_{IRS}}} \right),
	\end{split}
\end{equation}
\begin{equation}
	{f_2}\left( {\varphi } \right) = {\log _2}\varphi,
\end{equation}we can rewrite the optimization problem (21) as
\begin{align}
	\max\limits_{{\bf{W}}{\underline \succ}0,\varphi\ge1}\quad &{f_1}\left( {{\bf{W}},\eta } \right)-{f_2}\left( {\varphi } \right)\tag{24a}\\
	{\text{ s.t.}}\quad&(20b)-(20e).\tag{24b}
\end{align}

Considering that ${f_1}\left( {{\bf{W}},\eta } \right)$ and ${f_2}\left( {\varphi } \right)$ are two concave functions, the objective optimization function (24a) can be equivalently transformed into a convex function with the help of the D.C. programming approach \cite{12}. To solve the non-convex objective function, we apply the Taylor series expansion to approximate the concave function ${f_2}\left( {\varphi } \right)$ by a linear form. Supposing that $\bar \varphi $ is a feasible solution satisfying the constraints of the optimization problem (24), ${f_2}\left( {\varphi } \right)$ can be approximated by its first-order Taylor series expansion, i.e.,
\setcounter{equation}{24}
\begin{equation}
{f_2}\left( \varphi  \right) \le {f_2}\left( {\bar \varphi } \right) + \nabla {f_2}\left( {\bar \varphi } \right)\left( {\varphi  - \bar \varphi } \right),
\end{equation}where ${\nabla {f_2}\left( {\bar \varphi } \right)}$ is the gradient of the function ${f_2}\left( {\varphi } \right)$ with respect to $\varphi$ at point ${ {\bar \varphi } }$, given by
\begin{equation}
\nabla {f_2}\left( {\bar \varphi } \right) = 1/\left( {\bar \varphi \ln 2} \right).
\end{equation}
Then, by substituting (26) into (25), we can obtain
\begin{equation}
	\begin{split}
	{f_2}\left( \varphi  \right) \le {f_2}\left( {\bar \varphi } \right){\rm{ + }}\frac{{\varphi  - \bar \varphi }}{{\bar \varphi \ln 2}}.
	\end{split}
\end{equation}

Consequently, according to (27), the optimal solution to problem (24) can be achieved via the iterative process as follows
\begin{align}
	\left( {{{{\bf{ W}}}^{i}},{{\bar \varphi }^{i + 1}}} \right)& \notag\\
	=\max\limits_{{\bf{W}}{\underline \succ}0,\varphi\ge1}\quad &{f_1}\left( {{\bf{W}},\eta } \right) - {f_2}\left( {{{\bar \varphi }^i}} \right) - \frac{{\varphi  - {{\bar \varphi }^i}}}{{{{\bar \varphi }^i}\ln 2}} \tag{28a} \\
	{\text{ s.t.}}\quad&(20b)-(20e),\tag{28b}
\end{align}where $\left( {{{{\bf{W}}}^i},{{\bar \varphi }^{i+1}}} \right)$ represents the solution of the \textit{i}-th iteration. Now, the optimization problem (28) has satisfied the form of the convex optimization problem, whose optimal solution therefore can be obtained with the help of CVX.

\textit{Proposition 4}: The iterative solution process of problem (28) produces a sequence of better solutions that converges to the best solution of problem (24).

\quad\textit{Proof}: Appendix D.

By employing the \textit{Propositions} 3-4 and the D.C. approach, we propose an iterative algorithm for solving sub-problem 2 to obtain the optimal transmit beamforming at the CBS, summarized in Algorithm 2.

\begin{algorithm}[t]
	\caption{The Algorithm for Solving Sub-Problem 2}
	\LinesNumbered 
	\KwIn{${\bf{\Theta}}$, ${\bf{H}}_s$, ${\bf{H}}_p$, ${\bf{H}}_{Ek}$, $P_c^{max}$, $R_{sec}^{min}$, $I_p^{th}$}
	\KwOut{${\bf{W}}^*$}
	\SetKwProg{Fn}{Function}{}{end}
	\Fn{Outer\_Iteration}
	{
		Initialize $i=0$, $\eta^0=0$\;
		\ShowLn	
		\Repeat{$\left| {{\eta^{i}} - {\eta^{i-1}}} \right| \le \varepsilon $}{
			(\rmnum{1}) Call \textbf{Function} \textit{Inner\_Iteration} with $\eta^i$ to solve the optimal solution $({\bf{W}}^*, \varphi^*)=({\bf{W}}^n, \varphi^{n+1})$\;
			(\rmnum{2}) Update ${\eta^{i + 1}} := \frac{{{{\log }_2}\left( {1 + \frac{{tr\left( {{\bf{\Theta }}{{\bf{H}}_S}{{\bf{W}}^*}{{\bf{H}}_S^H}} \right)}}{{\sigma _S^2}}} \right) - {{\log }_2}{\varphi^* }}}{{\zeta tr\left( {{{\bf{W}}}^*} \right) + {P_{BS}} + {P_{IRS}}}}$\;
			(\rmnum{3}) Set $i:=i+1$\;
		}

		\ShowLn
		Obtain the maximum SEE $\eta^*=\eta^i$ and the optimal solution ${\bf{W}}^*$.   
	}	
	
	\SetKwProg{Fn}{Function}{}{end}
	\Fn{Inner\_Iteration($\eta$)}
	{
		\ShowLn
		Initialize $n=0$\;
		\ShowLn
		Find a feasible solution $\left( {{{{\bf{ W}}}^{0}},{{\varphi }^{1}}} \right)$ for problem (28) and calculate $f^0={f_1}\left( {{{\bf{W}}^0},\eta} \right) - {f_2}\left( {{{ \varphi }^{1)}}} \right)$\;
		\ShowLn
		\Repeat{$\left| {{f^{n}} - {f^{n-1}}} \right| \le \varepsilon$}
		{
			(i) Find the optimal solution $\left( {{{{\bf{ W}}}^{n+1}},{{\varphi }^{n+2}}} \right)$ of problem (28) for obtained $\left( {{{{\bf{ W}}}^{n}},{{\varphi }^{n+1}}} \right)$ by using CVX\;
			(ii) Compute $f^{n+1}:={f_{1}}\left( {{{\bf{W}}^{n+1}},\eta} \right) - {f_2}\left( {{{ \varphi }^{n+2}}} \right)$\;
			(iii) Set $n:=n+1$\;
		}
	}
\end{algorithm}

\subsection{The Overall Algorithm for Solving Problem (8)}
With the help of Charnes-Cooper transformation, penalty function, as well as D.C. programming, we propose an iterative alternating optimization algorithm to find the optimal solution $({\bf{w}^*}, {\bf{q}^*})$ of problem (8) among feasible solutions. Algorithm 3 illustrates the flow of the overall algorithm.

\textit{Remark 1:} In the scenario of non-cooperative Eves, the SR of SU can be written as ${R_{\sec }} = {\log _2}\left( {1 + \frac{{tr\left( {{\bf{\Theta }}{{\bf{H}}_S}{\bf{WH}}_S^H} \right)}}{{\sigma _S^2}}} \right) - \mathop {\max }\limits_{k \in \left\{ {1, \cdots ,K} \right\}} {\log _2}\left( {1 + \frac{{tr\left( {{\bf{\Theta }}{{\bf{H}}_{Ek}}{\bf{WH}}_{Ek}^H} \right)}}{{\sigma _{Ek}^2}}} \right)$,  where the key solution process of the SEE maximization problem remains the same as that in the scenario of cooperative Eves. Thus, our proposed algorithm can be easily extended to the scenario of non-cooperative Eves.

\begin{algorithm} [t]
	\caption{Iterative Alternating Optimization Algorithm for Solving Problem (8)}		
	\LinesNumbered 
	\KwIn{${\bf{H}}_s$, ${\bf{H}}_p$, ${\bf{H}}_{Ek}$, $P_c^{max}$, $R_{sec}^{min}$, $I_p^{th}$}
	\KwOut{${\bf{q}}^*$, ${\bf{w}}^*$}
	Initialize ${\bf{W}}^*=\sqrt {P_c^{\max }} \frac{{{\bf{h}}_{CS}^{}{\bf{h}}_{CS}^H}}{{{{\left\| {{\bf{h}}_{CS}^{}} \right\|}^2}}}, j=0, SEE^0=0$\;
	\ShowLn
	\Repeat{$\left| {{SEE^{j}} - {SEE^{j-1}}} \right| \le \varepsilon$} 
	{
		Set ${\bf{W}}$:=${\bf{W}}^*$\;
		Perform {\bf{Algorithm 1}} with given ${\bf{W}}$ to obtain ${\bf{\Theta}}^*$\;
		Set ${\bf{\Theta}}$:=${\bf{\Theta}}^*$\;
		Perform {\bf{Algorithm 2}} with given ${\bf{\Theta}}$ to obtain ${\bf{W}}^*$\;
		Update $SE{E^{j + 1}} := \frac{{{{\log }_2}\left( {1 + \frac{{tr\left( {{{\bf{\Theta }}^*}{{\bf{H}}_S}{{\bf{W}}^*}{\bf{H}}_S^H} \right)}}{{\sigma _S^2}}} \right) - {{\log }_2}\left( {1+\sum\limits_{k = 1}^K {\frac{{tr\left( {{{\bf{\Theta }}^*}{{\bf{H}}_{Ek}}{{\bf{W}}^*}{\bf{H}}_{Ek}^H} \right)}}{{\sigma _{Ek}^2}}} } \right)}}{{\zeta tr\left( {{{\bf{W}}^*}} \right) + {P_{BS}} + {P_{IRS}}}}$\;
		Set $j:=j+1$\;	
	}
	Obtain \\
	(i) the optimal transmit beamforming ${{\bf{w}^*}}$ through eigenvalue decomposition over ${\bf{W}}^*$\;
	(ii) the optimal reflect beamforming ${\bf{q}}^{*}$ through Cholesky decomposition \cite{a28} over ${\bf{\Theta}}^*$\;
	(iii) the maximum SEE, i.e., $SEE^j$.
\end{algorithm}

\section{A Low-Complexity Method for Sub-Problem 2} 
Note that the objective optimization function (28a) is in the logarithmic form, which means that the optimization problem (28) is nonlinear. As we know, the time of solving the nonlinear convex optimization problem is longer than that of the linear convex optimization problem. Considering this issue, we herein present an approximate method to convert (28) into a SOCP form. With the help of an auxiliary variable $s$, we reformulate problem (28) as
\begin{align}
	\left( {{{{\bf{ W}}}^{i}},{{\bar \varphi }^{i + 1}}} \right)& \notag\\
	=\max\limits_{{\bf{W}}{\underline \succ}0,\varphi\ge1}\quad &s - \eta \left( {\zeta tr\left( {\bf{W}} \right) + {P_{CBS}} + {P_{IRS}}} \right)\notag \\
	&\quad\quad- {\log _2}\left( {{{\bar \varphi }^i}} \right) - \frac{{\varphi  - {{\bar \varphi }^i}}}{{{{\bar \varphi }^i}\ln 2}}\tag{29a}\\
	{\text{ s.t.}}\quad&\ln \left( {1 + \frac{{tr\left( {{\bf{\Theta }}{{\bf{H}}_S}{\bf{WH}}_S^H} \right)}}{{\sigma _S^2}}} \right) \ge s\ln 2,	\tag{29b}\\
	&(20b)-(20e).\tag{29c}
\end{align}

By employing the identity given in \cite{13}, (29b) can be approximated with a series of second order cone constraints, namely,
\setcounter{equation}{29}
\begin{equation}
	\begin{array}{l}
		1 + {k_1} \ge \left\| {1 - {k_1},2 + s\ln 2/{2^{M + 1}}} \right\|,\\
		1 + {k_2} \ge \left\| {1 - {k_2},5/3 + s\ln 2/{2^M}} \right\|,\\
		1 + {k_3} \ge \left\| {1 - {k_3},2{k_1}} \right\|,\\
		{k_4} \ge {k_2} + {k_3}/24 + 19/72,\\
		1 + {k_m} \ge \left\| {1 - {k_m},2{k_{m - 1}}} \right\|,m = 5, \cdots ,M + 3,\\
		1 + {k_{M + 4}} \ge \left\| {1 - {k_{M + 4}},2{k_{M + 3}}} \right\|,\\
		1 + \frac{{tr\left( {{\bf{\Theta }}{{\bf{H}}_S}{\bf{WH}}_S^H} \right)}}{{\sigma _S^2}} \ge {k_{M + 4}},
	\end{array}
\end{equation}where ${k_m}(m = 1, \cdots ,M + 4)$ are auxiliary variables. $M$ represents the approximation accuracy.

By replacing (29b) with (30), we can convert problem (29) into a SOCP form approximately, which can be written as
\begin{align}
	\left( {{{{\bf{W}}}^{i }},{{\bar \varphi }^{i + 1}}} \right)& \notag\\
	=\max\limits_{{\bf{W}}{\underline \succ}0,\varphi\ge1}\quad &s - \eta \left( {\zeta tr\left( {\bf{W}} \right) + {P_{BS}} + {P_{IRS}}} \right)\notag \\
	&\quad\quad- {\log _2}\left( {{{\bar \varphi }^i}} \right) - \frac{{\varphi  - {{\bar \varphi }^i}}}{{{{\bar \varphi }^i}\ln 2}}\tag{31a} \\
	{\text{ s.t.}}\quad&(20b)-(20e), (30).\tag{31b}
\end{align}
Different from the nonlinear optimization problem (28), the optimization problem (31) is linear, which is beneficial for reducing the time of solving the optimization problem.

\section{Simulation Results}
In this section, simulation results are provided to validate the effectiveness of our proposed algorithm. 
We consider the simulation scenario as shown in Fig. 2: the PU, CBS, IRS and SU are located at (0, 0), (50, 0), ($x_{IRS}$, 10) and (100, 0) in meters, respectively. Also, the $K$ Eves are positioned uniformly along the line from (80, 0) to (90, 0) in meters. The channel coefficients are generated according to ${{\bf{h}}_{mn}} = \sqrt {{G_0}d_{mn}^{ - {c_{mn}}}} {{\bf{g}}_{mn}} (m \in \left\{ {C,I} \right\},n \in \left\{ {I,S,P,Ek} \right\},m \ne n)$, where $G_0=-30{\text{dB}}$ denotes the path loss (PL) of the reference point \cite{a9}.  ${\bf{g}}_{mn}$ and $c_{mn}$ stand for the Rayleigh channel fading and PL exponent between \textit{m} and \textit{n}, respectively. The PL exponents are set as $c_{CI}=c_{IS}=c_{IP}=c_{IEk}=2.2$, $c_{CS}=c_{CP}=c_{CEk}=3.75$ \cite{a7}. The other parameters are set as $N=4$, $L=60$, $K=2$, ${\sigma_S ^2}={\sigma_{Ek} ^2}=-100\text{dBm}$ \cite{re8}, $\zeta=1$, $P_{CBS}=23\text{dBm}$, $P_{IRS}=20\text{dBm}$, $I_p^{th}=7\text{dB}$ \cite{a32}, $R_{sec}^{min}=0.5\text{bit/s/Hz}$, $x_{IRS}=100$m and $\varepsilon  = {10^{ - 3}}$, unless otherwise stated. From Fig. 3 to Fig. 13, except that Fig. 4 is the result generated by single channel realization, all the other results are averaged over 200 channel realizations.

Fig. 3 shows the feasibility rate versus the maximum transmit power of CBS $P_c^{max}$ under different minimum acceptable SR thresholds $R_{sec}^{min}$. The feasibility rate is defined as the ratio of the number of feasible channel realizations to the total number of channel realizations, where the feasible channel realization means that there exists a feasible solution to the constrained problem with this channel realization. As we can see, with the increase of $P_c^{max}$, the feasibility rate of all algorithms also increases. In addition, with a given $P_c^{max}$, the increase of minimum acceptable SR threshold $R_{sec}^{min}$ will reduce the feasibility rate. These interesting phenomenons can be explained that small $P_c^{max}$ or large $R_{sec}^{min}$ may make QoS constraint unsatisfied, thereby reducing the feasibility rate. When $R_{sec}^{min}$ is 0.5bit/s/Hz, our proposed SEE maximization algorithm can achieve a relatively high feasibility rate even with a small $P_c^{max}=12$dBm. When $R_{sec}^{min}$ becomes larger (e.g., 1bit/s/Hz), the feasibility rate of our proposed SEE maximization algorithm under small $P_c^{max}$ reduces significantly. In this case, $P_c^{max}$ has to be greater than 16dBm to ensure an acceptable feasibility rate. Furthermore, we can find that the feasibility rate of SR maximization algorithm is the highest among all algorithms under the same condition. This is because that the QoS constraint and objective optimization function of the SR maximization algorithm have a consistent optimization trend, meaning that the algorithm has fewer infeasible areas, resulting in a higher feasibility rate. Compared with other benchmarks, our proposed SEE maximization algorithm can achieve a high feasibility rate, which is a bit smaller than that of SR maximization algorithm. However, our proposed algorithm can achieve a good trade-off between SR and power consumption.

\begin{figure}[t]
	\centering
	\includegraphics[width=9cm]{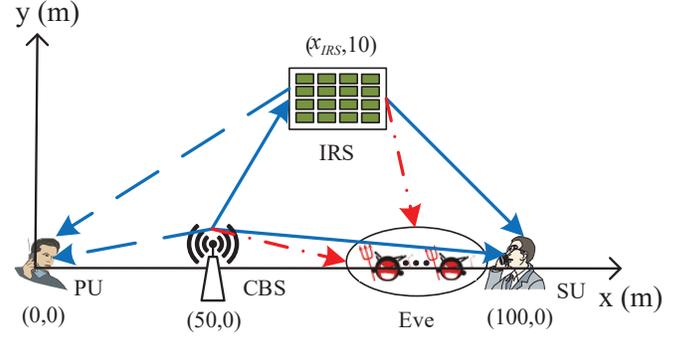}
	\caption{The simulation scenario.}
	\label{fig1}
\end{figure}
\begin{figure}[t]
	\centering
	\includegraphics[width=9cm]{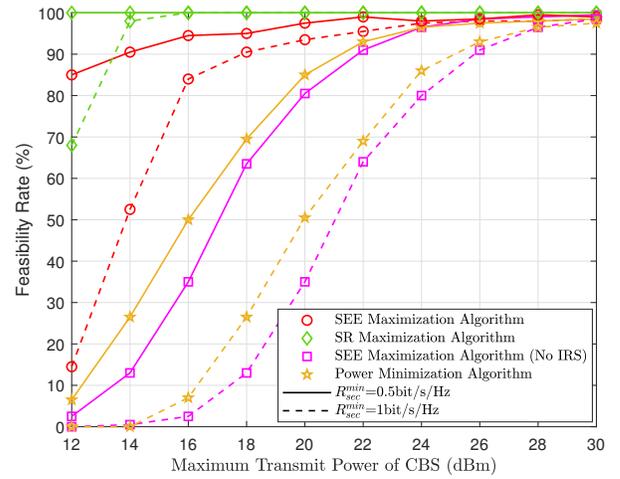}
	\caption{Feasibility rate.}
	\label{fig1}
\end{figure} 

Fig. 4 presents the convergence of our proposed algorithm when $P_c^{max}$=30dBm. As we can see, the SEE solved from the approximate SOCP form (31) has a perfect match with the SEE solved from (28), verifying the good precision of the SOCP approach. In addition, SEE increases with the iteration number, and finally reaches a stable value. It is shown that SEE converges to the optimal SEE within 15 iterations and good SEE performance can be achieved with only 4 iteration rounds, which demonstrates the effectiveness of our proposed algorithm. Moreover, a larger converged SEE value is reached with larger IRS elements, which can be explained that the IRS tends to reflect the signal stronger in the expected direction with more IRS elements. However, since the optimization variables increase with the increase of $L$, more IRS elements will bring a heavier computation burden, which is demonstrated in the form of a slower convergence speed with more phase shifts. Furthermore, we can find that when $I_p^{th}$ grows from -7dB to 7dB, the SEE increases. This can be explained that PU has the ability to bear larger interference from CBS for a larger $I_p^{th}$, meaning that the beamforming design of CBS and IRS has a greater degree of freedom to improve the SEE performance.

\begin{figure}[t]
	\centering
	\includegraphics[width=9cm]{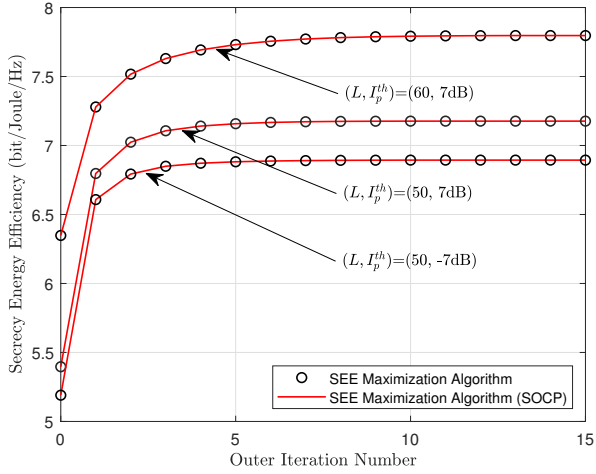}
	\caption{Convergence of the proposed algorithm.}
	\label{fig1}
\end{figure}
\begin{figure}[t]
	\centering
	\includegraphics[width=9cm]{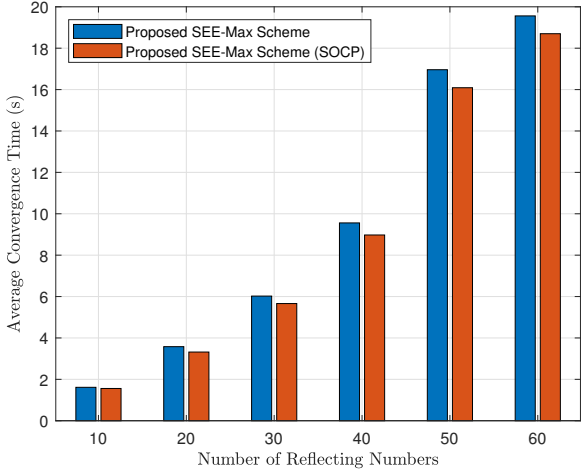}
	\caption{Average convergence time of the proposed algorithm.}
	\label{fig1}
\end{figure}

Fig. 5 investigates the average convergence time of our proposed algorithm versus the number of reflecting numbers $L$ with $P_c^{max}=12$dBm. As we can see, the real average time of convergence is no more than 20 seconds under small maximum transmit power of CBS $P_c^{max}$. In addition, as $L$ increases, the time of reaching convergence also increases. The reason is that the increasing $L$ leads to the expansion of selection range for optimization, which results in a longer time to search for the optimal solution. What's more, the average convergence time of the proposed approximate SOCP is less, which lies in that the time consumed to solve the linear convex optimization problem is shorter than that of the nonlinear convex optimization problem.


\begin{figure}[t]
	\centering
	\includegraphics[width=9cm]{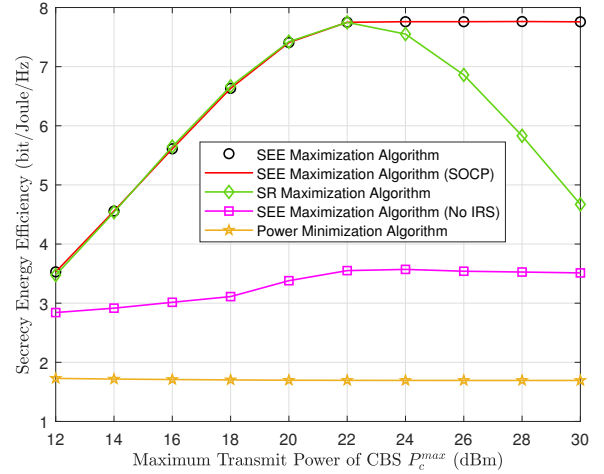}
	\caption{SEE versus CBS's maximum transmit power $P_c^{max}$.}
	\label{fig1}
\end{figure}


Fig. 6 presents the SEE versus CBS's maximum transmit power $P_c^{max}$. As we can see, there is almost no difference between our proposed SEE maximization algorithm and SR maximization scheme in terms of SEE performance when $12\text{dBm} \le P_c^{max}\le22\text{dBm}$. This is because that both of the algorithms try to tansmit the signal at the maximum transmit power provided by CBS. Nevertheless, when $P_c^{max}$ becomes larger further, our proposed SEE maximization algorithm only employs part of $P_c^{max}$ for transmission to ensure the maximum SEE to achieve a good trade-off between the SR and power consumption, while the SR maximization scheme continues to adopt all power provided by CBS to ensure the maximum SR, but ignores the reduction of SEE. Furthermore, we can find that the IRS-assisted schemes above are better than that without IRS, which verifies the effectiveness of the IRS introduced in the CRNs. This is because that the LoS path can be superimposed with the reflection signal with the same phase to enhance the signal, and with the opposite phase to weaken the signal in the expected directions. The SEE of the power minimization algorithm stays least. This can be interpreted as follows: the goal of the power minimization algorithm is to make the power consumption minimization. Therefore, in order to save energy, the transmit power only needs to meet the QoS requirements, i.e., to keep the SR at the minimum SR threshold $R_{sec}^{min}$. Under this circumstance, the SEE and SR are both relatively small.

\begin{figure}[t]
	\centering
	\includegraphics[width=9cm]{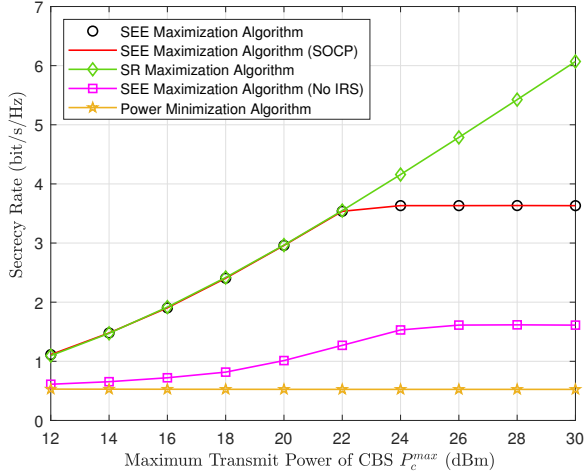}
	\caption{SR versus CBS's maximum transmit power $P_c^{max}$.}
	\label{fig1}
\end{figure}

\begin{figure}[t]
	\centering
	\includegraphics[width=9cm]{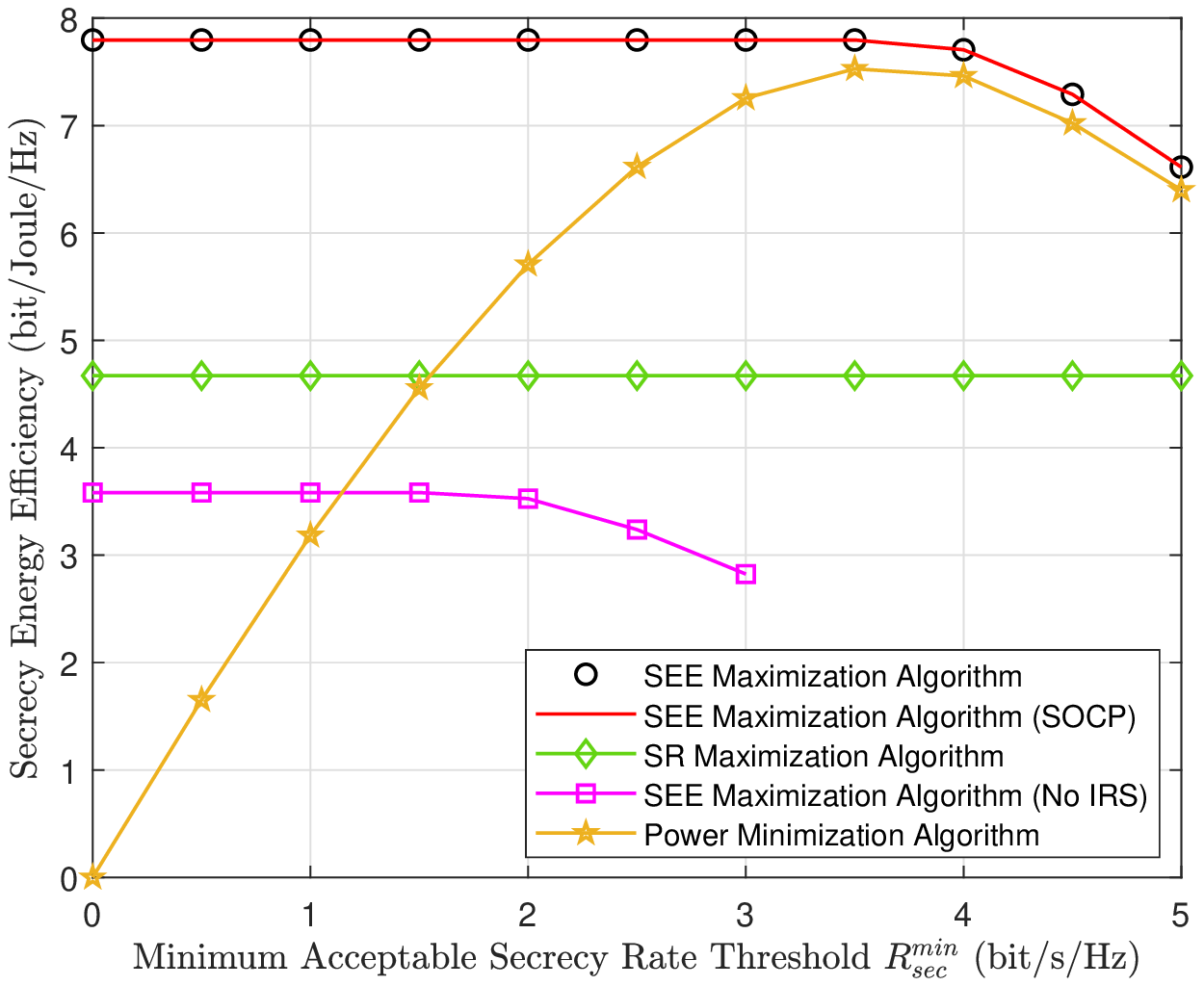}
	\caption{SEE versus SU's minimum acceptable SR threshold $R_{sec}^{min}$.}
	\label{fig1}
\end{figure}

\begin{figure}[t]
	\centering
	\includegraphics[width=9cm]{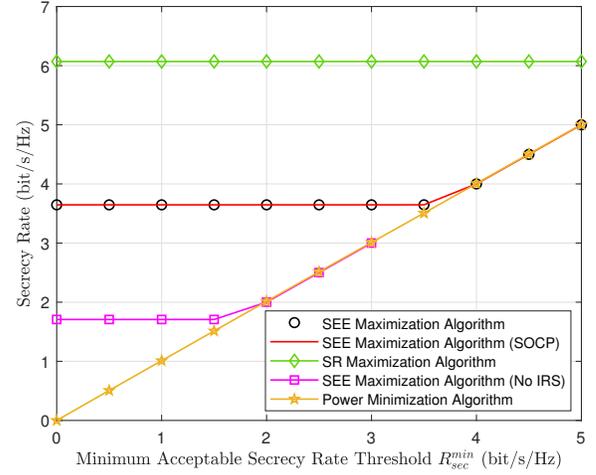}
	\caption{SR versus SU's minimum acceptable SR threshold $R_{sec}^{min}$.}
	\label{fig1}
\end{figure}

Fig. 7 depicts the SR versus CBS's maximum transmit power $P_c^{max}$. It is demonstrated that our proposed SEE maximization algorithm can achieve a comparable SR performance with SR maximization scheme when $P_c^{max}\le22$dBm. As $P_c^{max}$ grows further, the SR maximization scheme still adopts all power to ensure that SR is still at its maximum. However, our proposed algorithm can realize the trade-off between the SR and power consumption at the sacrifice of a little SR performance, which is in line with the idea of green communication in 6G. In addition, it is easy to find that the SR performance of the proposed algorithm is far better than that of the SEE maximization scheme without IRS, confirming the effectiveness of IRS in improving the SE and PLS in the CRNs. These curves show that the SEE maximization algorithm can realize a relatively high SR while ensuring SEE maximized. As for the power minimization method, the SR keeps at a lowest level, which owes to the fact that the purpose of this scheme is to make the power consumption minimization, causing the SR being the minimum SR threshold $R_{sec}^{min}$ to save power.

\begin{figure}[h]
	\centering
	\includegraphics[width=9cm]{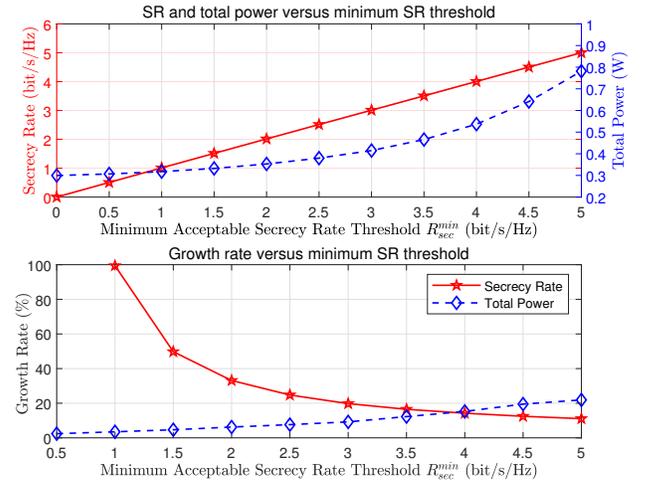}
	\caption{The curves of SR and total power.}
	\label{fig1}
\end{figure}
\begin{figure}[t]
	\centering
	\includegraphics[width=9cm]{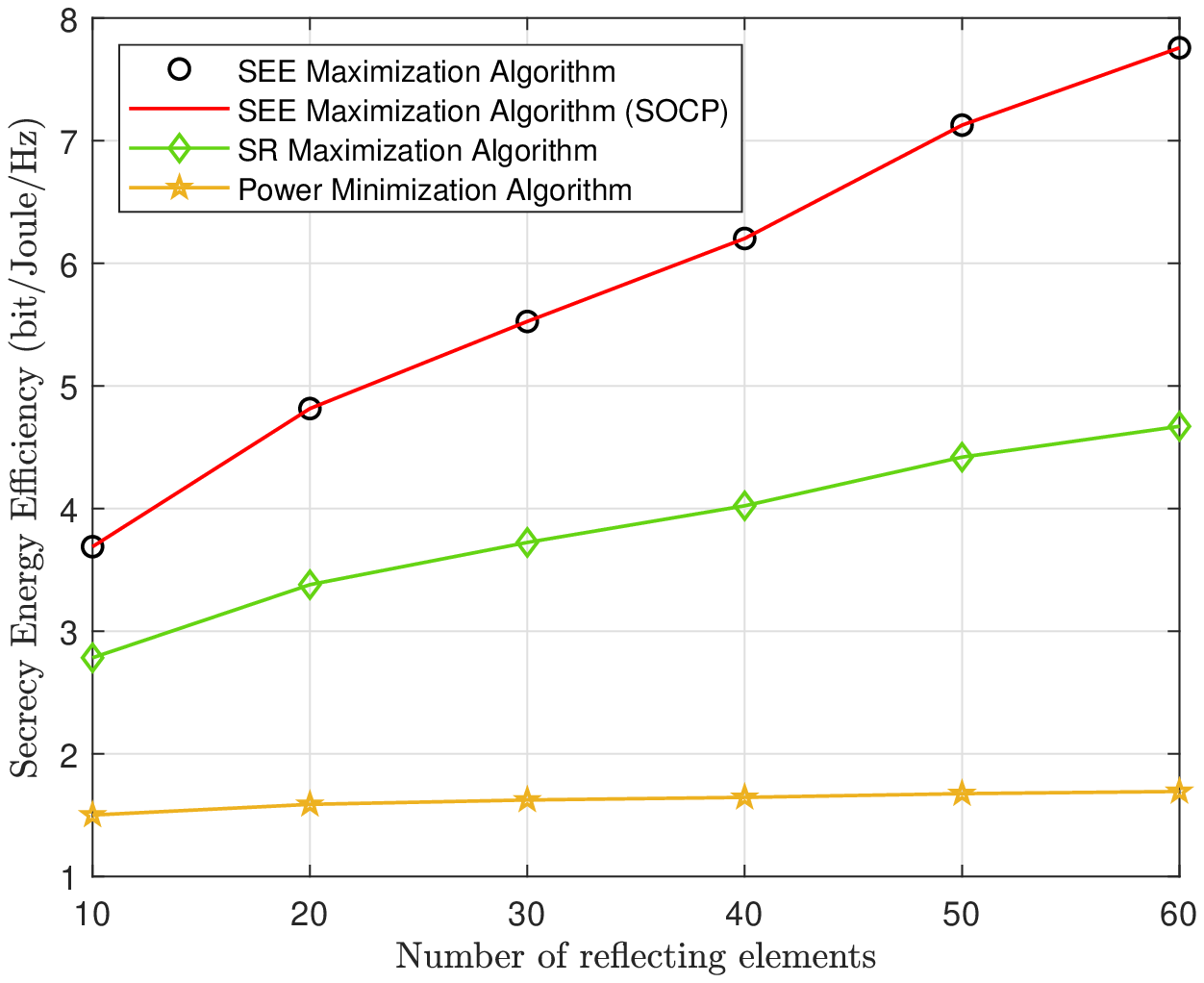}
	\caption{SEE versus the number of reflecting elements $L$.}
	\label{fig1}
\end{figure}
\begin{figure}[t]
	\centering
	\includegraphics[width=9cm]{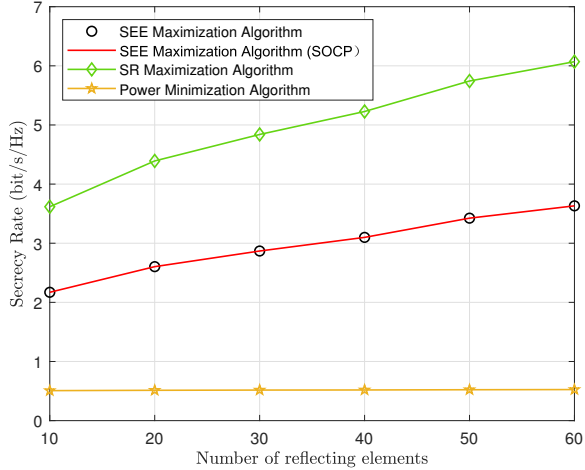}
	\caption{SR versus the number of reflecting elements $L$.}
	\label{fig1}
\end{figure}

Fig. 8 and Fig. 9 depict the SEE and SR versus SU's minimum acceptable SR threshold with $P_c^{max}=30$dBm, respectively. It can be seen from Fig. 8 and Fig. 9 that there exists a critical value for the minimum acceptable SR threshold $R_{sec}^{min}$ of our proposed SEE maximization algorithm. Before reaching this critical value, with the increase of the SR threshold, the SEE and SR of our proposed algorithm remain unchanged, and the SR is equal to the critical value when the SR threshold $R_{sec}^{min}$ is just the critical vale. After reaching the critical SR threshold, increasing the SR threshold will reduce the SEE, because the SR of the network will increase synchronously with the increase of the SR threshold, which requires a large amount of improvement of CBS's transmit power to meet the requirements of the increased SR. This shows that the SEE maximization algorithm proposed in this paper can maximize the SEE of the network while also maximizing the SR of the network under the condition of meeting the minimum acceptable SR constraint. The same analysis also applies to the SEE maximization algorithm without IRS. The SEE of the SR maximization algorithm remains unchanged. Since the CBS under the SR maximization scheme consumes almost all power to ensure that SR reaches the maximum value no matter what the value of the minimum SR threshold $R_{sec}^{min}$ is, the SR will remain unchanged and the SR achieved by this scheme is the highest among all the schemes. Due to the unchanged SR and the almost all available power consumed by CBS, the SEE of this scheme also keeps constant. Moreover, the SEE of the power minimization scheme first increases, and then begins to decline when $R_{sec}^{min}$ is greater than 3.5 bits/s/Hz. To explain this, we depict the SR and total power (TP) versus the minimum acceptable SR threshold for power minimization algorithm in Subfig. 1 of Fig. 10, where the red curve is based on the ordinate scale on the left, and the blue curve is based on the ordinate scale on the right. Based on Subfig. 1, we give the growth rate curves of SR and TP in Subfig. 2, defined as $\rm{GR}=\frac{{{\rm{S}}{{\rm{R}}_c}\left( {{\rm{T}}{{\rm{P}}_c}} \right) - {\rm{S}}{{\rm{R}}_p}\left( {{\rm{T}}{{\rm{P}}_p}} \right)}}{{{\rm{S}}{{\rm{R}}_p}\left( {{\rm{T}}{{\rm{P}}_p}} \right)}} \times 100\% $, where $\rm{GR}$, ${{\rm{S}}{{\rm{R}}_c}\left( {{\rm{T}}{{\rm{P}}_c}} \right)}$ and ${{\rm{S}}{{\rm{R}}_p}\left( {{\rm{T}}{{\rm{P}}_p}} \right)}$ represent the growth rate, the current SR (TP) value, and the previous SR (TP) value, respectively. Obviously, when the SR threshold is no more than 3.5 bits/s/Hz, the growth rate of SR is larger than that of TP, that is to say, ${\rm{SEE}} = \frac{{{\rm{SR}}}}{{{\rm{TP}}}}$ will show an increasing trend. Under this circumstance, the increase of SR will make a greater contribution to the improvement of SEE performance. However, when the SR threshold is larger than 3.5 bits/s/Hz, the growth rate of SR is smaller than that of TP, which means that ${\rm{SEE}} = \frac{{{\rm{SR}}}}{{{\rm{TP}}}}$ will show a downward trend. In this case, even if the SR is large, the contribution to the SEE performance is far less than the impact of the large TP, resulting in a decrease in SEE performance. The analysis above also shows that there is a trade-off between the SR and power consumption. Additionally, since the power minimization method aims to minimize the transmit power, the SR of the method is definitely the SR threshold, without wasting any power to increase the SR, which is coincide with the observation in Fig. 9 that the SR of the power minimization method increases with the minimum SR threshold linearly.

Fig. 11 and Fig. 12 depict the SEE and SR versus the number of reflecting elements $L$ with $P_c^{max}=30$dBm. Obviously, the SEE of our proposed SEE maximization algorithm significantly improves as $L$ increases. This can be explained that a greater number of reflecting elements can enhance the desired signals for SU more flexibly, which further achieves a better SEE performance. Moreover, the SEE increasing gain of our proposed algorithm is higher than those obtained by other benchmarks, indicating that our proposed algorithm can effectively exploit IRS to assist the secure transmission. In addition, we can find that although our proposed algorithm can guarantee the secure transmission requirement of the system, the SR of our proposed algorithm is lower than that of SR maximization algorithm. Since the maximum transmit power of CBS $P_c^{max}$ is large, our proposed SEE maximization algorithm will sacrifice some SR performance in exchange for SEE performance, thereby ensuring a trade-off between SR and power consumption.

\begin{figure}[t]
	\centering
	\includegraphics[width=9cm]{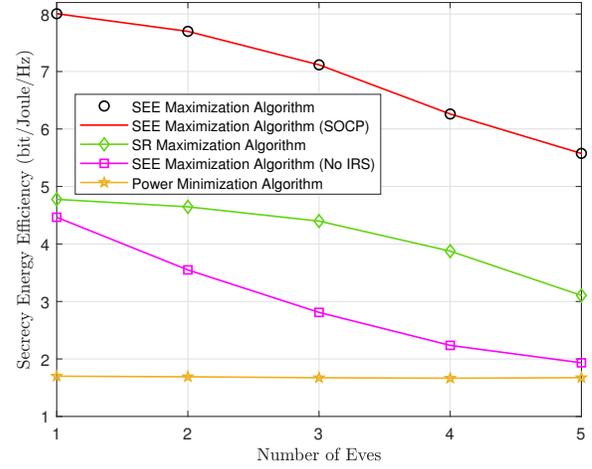}
	\caption{SEE versus the number of Eves $K$.}
	\label{fig1}
\end{figure}

Fig. 13 demonstrates the SEE versus the number of Eve with $P_c^{max}=30$dBm. With the increase of the number of wiretap links, the SEE obtained by the proposed algorithm is higher than other benchmarks, which shows that the proposed algorithm can effectively achieve a good trade-off between the secure transmission and power consumption. Moreover, as the number of Eve increases, the SEE of the proposed algorithm shows a downward trend. The result is reasonable since with the restriction of secrecy rate constraint, the increase of Eve number will reduce the degree of freedom of CBS and IRS beamforming vectors, resulting in a decrease of SR, which further reduces the SEE performance.

\begin{figure}[t]
	\centering
	\includegraphics[width=9cm]{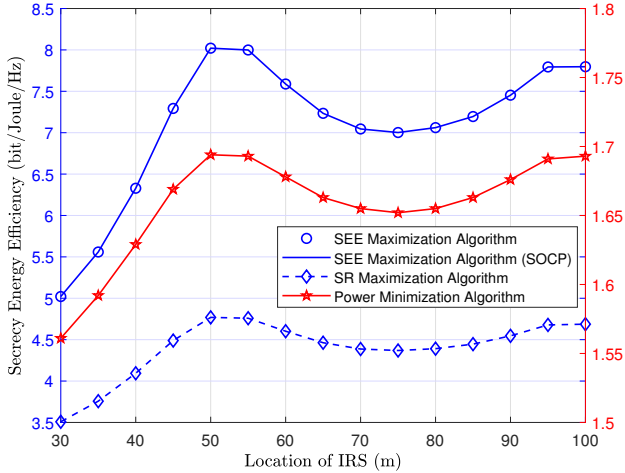}
	\caption{SEE versus the location of IRS.}
	\label{fig1}
\end{figure}

Fig. 14 depicts the SEE versus the location of IRS with $P_c^{max}=30$dBm, where the blue curves are based on the ordinate scale on the left, and the red curves are based on the ordinate scale on the right. It is shown that there are two optimal locations in this figure, i.e., $x_{IRS}=50$m and $x_{IRS}=100$m, which are the optimal locations in both the proposed algorithm and the benchmarks. Furthermore, the IRS should not be deployed in the middle location between the CBS and SU. Actually, the worst case occurs when the IRS is far away from SU and close to the PU. These results indicate that, in order to obtain a higher SEE, the IRS should be deployed in the vicinity of the CBS or SU.

\section{Conclusion}

This paper has studied an IRS-assisted CRN to improve SU's SEE. To this end, we investigated a SEE maximization problem by jointly optimizing the transmit beamforming at the CBS and reflect beamforming at the IRS under the premise that certain SR of SU and power of cognitive transmission are guaranteed, and the leakage interference to PU restricted under a predefined threshold. To deal with the complicated non-convex form, we proposed an iterative alternating optimization algorithm by decoupling the original problem into two sub-problems. Besides, for the purpose of decreasing the computational complexity, we provided a SOCP approximation approach. The simulation results showed that IRS can help significantly improve the SE and enhance the PLS in the CRNs and demonstrated that the proposed algorithm can achieve the highest SEE among all the benchmark methods, indicating that our algorithm can achieve a good trade-off between the SR and energy consumption in the IRS-assisted CRNs. In addition, it is interesting to find that there exists a critical value for the minimum acceptable SR threshold of our proposed algorithm, which further indicates that both the SEE and SR can be maximized under the condition that the minimum acceptable SR constraint is satisfied. Our proposed algorithm can be extended to the case with multiple SUs and PUs when the CBS transmits the same signal to the SUs. However, when the CBS transmits different signals to corresponding SU, there will exist inter-user interference, which will be the content of our follow-up research.

\appendices
\section{Proof of Proposition 1}
Rewrite the problem (11) as
\begin{align}
	\min\limits_{{\bf{\Theta}}}\quad &\frac{{1 + \sum\limits_{k = 1}^K {\frac{{tr\left( {{\bf{\Theta }}{{\bf{H}}_{Ek}}{\bf{WH}}_{Ek}^H} \right)}}{{\sigma _{Ek}^2}}} }}{{1 + \frac{{tr\left( {{\bf{\Theta }}{{\bf{H}}_S}{\bf{WH}}_S^H} \right)}}{{\sigma _S^2}}}}\tag{32a}\\
	{\text{ s.t.}}\quad
	&tr\left( {{\bf{\Theta }}{{\bf{H}}_P}{\bf{WH}}_P^H} \right) \le I_p^{th}, \tag{32b}\\
	&{\bf{\Theta }}{\underline \succ}0,\quad rank({\bf{\Theta}})=1,\quad{\rm{  and  }}\quad{\left[ {\bf{\Theta }} \right]_{l, l}} = 1, \forall l \in {\mathbb L}.\tag{32c}
\end{align}	
With the help of the auxiliary variable ${\bf{A}}=a{\bf{\Theta}}$, $a>0$, (32) can be further reformulated as
\begin{align}
	\min\limits_{{\bf{A}}{\underline \succ}0, a\ge0}\quad &\frac{{a + \sum\limits_{k = 1}^K {\frac{{tr\left( {{\bf{A }}{{\bf{H}}_{Ek}}{\bf{WH}}_{Ek}^H} \right)}}{{\sigma _{Ek}^2}}} }}{{a + \frac{{tr\left( {{\bf{A }}{{\bf{H}}_S}{\bf{WH}}_S^H} \right)}}{{\sigma _S^2}}}}\tag{33a}\\
	{\text{ s.t.}}\quad
	&tr\left( {{\bf{A }}{{\bf{H}}_P}{\bf{WH}}_P^H} \right) \le aI_p^{th}, \tag{33b}\\
	&rank({\bf{A}})=1,\quad{\rm{  and  }}\quad{\left[ {\bf{A }} \right]_{l, l}} = a, \forall l \in {\mathbb L},\tag{33c}
\end{align}
which can be equivalently written as
\begin{align}
	{\mathop {\min }\limits_{{\bf{A}}{\underline \succ}0, a\ge0} \quad }&{a + \sum\limits_{k = 1}^K {\frac{{tr\left( {{\bf{A}}{{\bf{H}}_{Ek}}{\bf{WH}}_{Ek}^H} \right)}}{{\sigma _{Ek}^2}}} }\tag{34a}\\
	{{\rm{ s}}{\rm{.t}}{\rm{.}}\quad }&{a + \frac{{tr\left( {{\bf{A}}{{\bf{H}}_S}{\bf{WH}}_S^H} \right)}}{{\sigma _S^2}} = 1},\tag{34b}\\
	&tr\left( {{\bf{A }}{{\bf{H}}_P}{\bf{WH}}_P^H} \right) \le aI_p^{th}, \tag{34c}\\
	{}&{ rank({\bf{A}}) = 1,\quad {\rm{and}}\quad {{\left[ {\bf{A}} \right]}_{l,l}} = a,\forall l \in {\mathbb L} }.\tag{34d}
\end{align}
Rewriting (34b) as ${a + \frac{{tr\left( {{\bf{A}}{{\bf{H}}_S}{\bf{WH}}_S^H} \right)}}{{\sigma _S^2}} \ge 1}$ does not change the optimal solution of (34), which can be explained as follows: assume that $\left( {{{\bf{A}}^*},{a^*}} \right)$ is the optimal solution satisfying ${a^* + \frac{{tr\left( {{\bf{A}^*}{{\bf{H}}_S}{\bf{WH}}_S^H} \right)}}{{\sigma _S^2}} > 1}$. Then, there definitely exists a certain vale $0<\beta<1$, enabling us to choose a feasible point 
$\left( {{\bf{\bar A}},\bar a} \right) = \left( {\beta {{\bf{A}}^*},\beta {a^*}} \right)$ to make $\frac{{tr\left( {{\bf{\bar A}}{{\bf{H}}_S}{\bf{WH}}_S^H} \right)}}{{\sigma _S^2}} + \bar a = 1$. Obviously, $\left( {{\bf{\bar A}},\bar a} \right)$ meets the constraints (34c)-(34d) and $\left( {{\bf{\bar A}},\bar a} \right)$ can be proved to provide a smaller optimization value (34a) than that provided from $\left( {{{\bf{A}}^*},{a^*}} \right)$, which is in contradiction with the assumption that $\left( {{{\bf{A}}^*},{a^*}} \right)$ is the optimal solution. Therefore, we can rewrite the constraint (34b) as ${a + \frac{{tr\left( {{\bf{A}}{{\bf{H}}_S}{\bf{WH}}_S^H} \right)}}{{\sigma _S^2}} \ge 1}$, which is a convex constraint. Based on the analysis above, (34) can be further expressed as
\begin{align}
	{\mathop {\min }\limits_{{\bf{A}}{\underline \succ}0, a\ge0} \quad }&{a + \sum\limits_{k = 1}^K {\frac{{tr\left( {{\bf{A}}{{\bf{H}}_{Ek}}{\bf{WH}}_{Ek}^H} \right)}}{{\sigma _{Ek}^2}}} }\tag{35a}\\
	{{\rm{ s}}{\rm{.t}}{\rm{.}}\quad }&{a + \frac{{tr\left( {{\bf{A}}{{\bf{H}}_S}{\bf{WH}}_S^H} \right)}}{{\sigma _S^2}} \ge 1},\tag{35b}\\
	&tr\left( {{\bf{A }}{{\bf{H}}_P}{\bf{WH}}_P^H} \right) \le aI_p^{th}, \tag{35c}\\
	{}&{ rank({\bf{A}}) = 1,\quad {\rm{and}}\quad {{\left[ {\bf{A}} \right]}_{l,l}} = a,\forall l \in {\mathbb L} }.\tag{35d}
\end{align}

To simplify the objective function, we introduce the auxiliary variable $t$. Then, (35) can be further written as
\begin{align}
	{\mathop {\min }\limits_{{\bf{A}}{\underline \succ}0, a\ge0} \quad }&{t }\tag{36a}\\
	{{\rm{ s}}{\rm{.t}}{\rm{.}}\quad }&{a + \frac{{tr\left( {{\bf{A}}{{\bf{H}}_S}{\bf{WH}}_S^H} \right)}}{{\sigma _S^2}} \ge 1},\tag{36b}\\
	&a + \sum\limits_{k = 1}^K {\frac{{tr\left( {{\bf{A}}{{\bf{H}}_{Ek}}{\bf{WH}}_{Ek}^H} \right)}}{{\sigma _{Ek}^2}}}  = t,\tag{36c}\\
	&tr\left( {{\bf{A }}{{\bf{H}}_P}{\bf{WH}}_P^H} \right) \le aI_p^{th}, \tag{36d}\\
	{}&{rank({\bf{A}}) = 1,\quad {\rm{and}}\quad {{\left[ {\bf{A}} \right]}_{l,l}} = a,\forall l \in {\mathbb L} }.\tag{36e}
\end{align}Noting that the objective function is to minimize $t$, we can rewrite the non-convex constraint (36c) as a convex constraint $a + \sum\limits_{k = 1}^K {\frac{{tr\left( {{\bf{A}}{{\bf{H}}_{Ek}}{\bf{WH}}_{Ek}^H} \right)}}{{\sigma _{Ek}^2}}}  \le t$, which will meet the equality constraint (36c) when the optimal solution of (36) are obtained. Therefore, we can formulate the problem (36) as
\begin{align}
	{\mathop {\min }\limits_{{\bf{A}}{\underline \succ}0, a\ge0} \quad }&{t }\tag{37a}\\
	{{\rm{ s}}{\rm{.t}}{\rm{.}}\quad }&{a + \frac{{tr\left( {{\bf{A}}{{\bf{H}}_S}{\bf{WH}}_S^H} \right)}}{{\sigma _S^2}} \ge 1},\tag{37b}\\
	&a + \sum\limits_{k = 1}^K {\frac{{tr\left( {{\bf{A}}{{\bf{H}}_{Ek}}{\bf{WH}}_{Ek}^H} \right)}}{{\sigma _{Ek}^2}}}  \le t,\tag{37c}\\
	&tr\left( {{\bf{A }}{{\bf{H}}_P}{\bf{WH}}_P^H} \right) \le aI_p^{th}, \tag{37d}\\
	{}&{rank({\bf{A}}) = 1,\quad {\rm{and}}\quad {{\left[ {\bf{A}} \right]}_{l,l}} = a,\forall l \in {\mathbb L} }.\tag{37e}
\end{align}

\section{Proof of Proposition 2}
In order to prove the equivalence of problems (12) and (13), we just need to prove the correctness of $rank\left( {\bf{A}} \right) = 1 \Leftrightarrow tr\left( {\bf{A}} \right) - {\lambda _{\max }}\left( {\bf{A}} \right) \le 0$.

Firstly, we will prove $rank\left( {\bf{A}} \right) = 1 \Rightarrow  tr\left( {\bf{A}} \right) - {\lambda _{\max }}\left( {\bf{A}} \right) \le 0$. Since $rank\left( {\bf{A}} \right) = 1$, the rank of the column vector group of $\bf{A}$ is 1. Assume that the first column of $\bf{A}$ is ${\bf{x}}{\rm{ = }}{\left( {{x_{\rm{1}}},{x_2}, \cdots ,{x_L}} \right)^T} \ne 0\left( {{x_{\rm{1}}} \ne 0} \right)$, and the other columns can be represented by $\bf{x}$ linearly, so $\bf{A}$ can be represented as ${\bf{A}} = \left( {{y_1}{\bf{x}},{y_2}{\bf{x}}, \cdots ,{y_L}{\bf{x}}} \right) = {\bf{x}}{{\bf{y}}^T}$, where $y_1=1$, ${\bf{y}} = {\left( {{y_1},{y_2}, \cdots ,{y_L}} \right)^T}$. Then, the characteristic polynomials of the matrix $\bf{A}$ can be expressed as 
\setcounter{equation}{37}
\begin{equation}
\begin{array}{l}
	\left| {\lambda {{\bf{I}}_L} - {\bf{A}}} \right| = \left| {\begin{array}{*{20}{c}}
			{\lambda  - {x_{\rm{1}}}{y_{\rm{1}}}}&{ - {x_{\rm{1}}}{y_2}}& \cdots &{ - {x_{\rm{1}}}{y_L}}\\
			{ - {x_{\rm{2}}}{y_{\rm{1}}}}&{\lambda  - {x_2}{y_2}}& \cdots &{ - {x_2}{y_L}}\\
			\vdots & \vdots &{}& \vdots \\
			{ - {x_L}{y_{\rm{1}}}}&{ - {x_L}{y_2}}& \cdots &{\lambda  - {x_L}{y_L}}
	\end{array}} \right|\\
	\begin{array}{*{20}{c}}
		{{r_i} - \frac{{{x_2}}}{{{x_1}}}r_1}\\
		{\overline{\overline {i = 2, \cdots ,L}} }
	\end{array}\left| {\begin{array}{*{20}{c}}
			{\lambda  - {x_{\rm{1}}}{y_{\rm{1}}}}&{ - {x_{\rm{1}}}{y_2}}& \cdots &{ - {x_{\rm{1}}}{y_L}}\\
			{ - \frac{{{x_{\rm{2}}}}}{{{x_{\rm{1}}}}}{\lambda}}&\lambda & \cdots &{\rm{0}}\\
			\vdots & \vdots &{}& \vdots \\
			{ - \frac{{{x_L}}}{{{x_{\rm{1}}}}}{\lambda}}&{\rm{0}}& \cdots &\lambda 
	\end{array}} \right|\\
	\begin{array}{*{20}{c}}
		{\underline{\underline {{c_1} + \sum\limits_{i = 2}^L {\frac{{{x_i}}}{{{x_1}}}{c_i}} }} }\\
		{}
	\end{array}\left| {\begin{array}{*{20}{c}}
			{\lambda  - \sum\limits_{l = 1}^L {{x_l}{y_l}} }&{ - {x_{\rm{1}}}{y_2}}& \cdots &{ - {x_{\rm{1}}}{y_L}}\\
			0&\lambda & \cdots &{\rm{0}}\\
			\vdots & \vdots &{}& \vdots \\
			0&{\rm{0}}& \cdots &\lambda 
	\end{array}} \right|\\
	= {\lambda ^{L - 1}}\left( {\lambda  - \sum\limits_{l = 1}^L {{x_l}{y_l}} } \right),
\end{array}
\end{equation}
where $r_i$ and $c_i$ represent the \emph{i}-th row and column of the matrix ${\lambda {{\bf{I}}_L} - {\bf{A}}}$, respectively.
By letting $\left| {\lambda {{\bf{I}}_L} - {\bf{A}}} \right|$ be zero, we can obtain the eigenvalues of $\bf{A}$ as ${\lambda _1} = {\lambda _2} =  \cdots  = {\lambda _{L - 1}} = 0$, ${\lambda _L} = \sum\limits_{l = 1}^L {{x_l}{y_l}}$. Thus, we have
\begin{equation}
	\begin{split}
	&tr\left( {\bf{A}} \right) - {\lambda _{\max }}\left( {\bf{A}} \right) = \sum\limits_{l = 1}^L {{\lambda _l}}  - {\lambda _{\max }}\left( {\bf{A}} \right)\\
	&= {\lambda _L} - {\lambda _{\max }}\left( {\bf{A}} \right) = \left\{ {\begin{array}{*{20}{c}}
			{{\lambda _L},}&{{\lambda _L} < 0}\\
			{0,}&{{\lambda _L} \ge 0}
	\end{array}} \right.\\
	&\le 0.
    \end{split}
\end{equation}

Next, we will prove $tr\left( {\bf{A}} \right) - {\lambda _{\max }}\left( {\bf{A}} \right) \le 0 \Rightarrow rank\left( {\bf{A}} \right) = 1$. Due to the fact that $tr\left( {\bf{A}} \right) - {\lambda _{\max }}\left( {\bf{A}} \right) \ge 0$ always holds for any matrix ${\bf{A}}{\underline \succ}0$, $tr\left( {\bf{A}} \right) - {\lambda _{\max }}\left( {\bf{A}} \right) \le 0$ can be written as $tr\left( {\bf{A}} \right) - {\lambda _{\max }}\left( {\bf{A}} \right) = 0$ equivalently. Denote the eigenvalues of $\bf{A}({\bf{A}}{\underline \succ}0)$ as ${\lambda _1},{\lambda _2}, \cdots ,{\lambda _L}\left( {0\le{\lambda _1} \le {\lambda _2} \le  \cdots  \le {\lambda _L}} \right)$, and we have
\begin{equation}
	\begin{split}
	&tr\left( {\bf{A}} \right) - {\lambda _{\max }}\left( {\bf{A}} \right) = \sum\limits_{l = 1}^L {{\lambda _l}}  - {\lambda _{\max }}\left( {\bf{A}} \right) = 0\\
	&\Rightarrow \sum\limits_{l = 1}^L {{\lambda _l}}  = {\lambda _{\max }}\left( {\bf{A}} \right) \Rightarrow {\lambda _L} + \sum\limits_{l = 1}^{L - 1} {{\lambda _l}}  = {\lambda _L}\\
	&\Rightarrow \sum\limits_{l = 1}^{L - 1} {{\lambda _l}}  = 0.
    \end{split}
\end{equation}
Since ${{\lambda _l} \ge 0}$, we can further obtain ${\lambda _1} = {\lambda _2} =  \cdots  = {\lambda _{L - 1}} = 0$. If $\lambda_L=0$, the positive semi-definite matrix $\bf{A}$ will be a zero matrix, which cannot be an optimal solution. Thus, $\lambda_L$ must be positive, meaning that the positive semi-definite matrix $\bf{A}$ has one and only one non-zero eigenvalue, i.e. $rank({\bf{A}})=1$.

\section{Proof of Proposition 3}
Let ${\lambda _i} \ge 0\left( {i = 1,2,3,4} \right)$ and ${\bf{X}}$ be the dual variables of the constraints (20b)-(20e) and ${\bf{W}}$ respectively, then the Lagrangian function of the optimization problem (21) (ignoring the rank-1 constraint) can be expressed as
\begin{equation}
\begin{split}
	Lag =  &- {\log _2}\left( {1 + \frac{{tr\left( {{\bf{\Theta }}{{\bf{H}}_S}{\bf{WH}}_S^H} \right)}}{{\sigma _S^2}}} \right) + {\log _2}\varphi \\
	&+ \eta \left( {\zeta tr\left( {\bf{W}} \right) + {P_{CBS}} + {P_{IRS}}} \right)\\
	&+ {\lambda _1}\left( {\sum\limits_{k = 1}^K {\frac{{tr\left( {{\bf{\Theta }}{{\bf{H}}_{Ek}}{\bf{WH}}_{Ek}^H} \right)}}{{\sigma _{Ek}^2}}}  - \varphi  + 1} \right)\\
	&+ {\lambda _2}\left( {{2^{R_{\sec }^{\min }}}\varphi  - \frac{{tr\left( {{\bf{\Theta }}{{\bf{H}}_S}{\bf{WH}}_S^H} \right)}}{{\sigma _S^2}} - 1} \right)\\
	&+ {\lambda _3}\left( {tr\left( {{\bf{\Theta }}{{\bf{H}}_p}{\bf{WH}}_p^H} \right) - I_p^{th}} \right)\\
	&+ {\lambda _4}\left( {tr\left( {\bf{W}} \right) - P_c^{\max }} \right) - tr\left( {{\bf{XW}}} \right).
\end{split}
\end{equation}

According to KKT conditions, we have
\begin{equation}
	\begin{split}
	{{\bf{X}}^*} =  - \left( {\frac{1}{{\sigma _S^2 + tr\left( {{\bf{\Theta }}{{\bf{H}}_S}{{\bf{W}}^*}{\bf{H}}_S^H} \right)}} + \frac{{\lambda _2^*}}{{\sigma _S^2}}} \right){\bf{H}}_S^H{\bf{\Theta }}{{\bf{H}}_S}\\
	+ \lambda _1^*\sum\limits_{k = 1}^K {\frac{{{\bf{H}}_{Ek}^H{\bf{\Theta }}{{\bf{H}}_{Ek}}}}{{\sigma _{Ek}^2}}}  + \lambda _3^*{\bf{H}}_p^H{\bf{\Theta }}{{\bf{H}}_p} + \left( {\eta \zeta  + \lambda _4^*} \right){{\bf{I}}_N},
\end{split}
\end{equation}
\begin{equation}
{{\bf{X}}^*}{{\bf{W}}^*} = {\bf{0}},
\end{equation}
\begin{equation}
{{\bf{W}}^*} {\underline \succ} {\bf{0}},\quad\lambda _i^* \ge 0\left( {i = 1,2,3,4} \right),
\end{equation}
where ${{\bf{W}}^*}$, ${{\bf{X}}^*}$, and ${{{\lambda}}_i^*}$ are the optimal solutions of ${{\bf{W}}}$, ${{\bf{X}}}$, and ${{{\lambda}}}_i$, respectively.

By substituting (42) into (43), we can obtain
\begin{equation}
	\begin{split}
	&\left( {\lambda _1^*\sum\limits_{k = 1}^K {\frac{{{\bf{H}}_{Ek}^H{\bf{\Theta }}{{\bf{H}}_{Ek}}}}{{\sigma _{Ek}^2}}}  + \lambda _3^*{\bf{H}}_p^H{\bf{\Theta }}{{\bf{H}}_p} + \left( {\eta \zeta  + \lambda _4^*} \right){{\bf{I}}_N}} \right){{\bf{W}}^*}\\
	&= \left( {\frac{1}{{\sigma _S^2 + tr\left( {{\bf{\Theta }}{{\bf{H}}_S}{{\bf{W}}^*}{\bf{H}}_S^H} \right)}} + \frac{{\lambda _2^*}}{{\sigma _S^2}}} \right){\bf{H}}_S^H{\bf{\Theta }}{{\bf{H}}_S}{{\bf{W}}^*}.
\end{split}
\end{equation}
Since $\lambda _1^*,\lambda _3^*,\lambda _4^* \ge 0$, and ${\lambda _1^*\sum\limits_{k = 1}^K {\frac{{{\bf{H}}_{Ek}^H{\bf{\Theta }}{{\bf{H}}_{Ek}}}}{{\sigma _{Ek}^2}}}  + \lambda _3^*{\bf{H}}_p^H{\bf{\Theta }}{{\bf{H}}_p} + \left( {\eta \zeta  + \lambda _4^*} \right){{\bf{I}}_N}}$ is a full rank matrix, we can attain
\begin{equation}
	\begin{split}
	&rank\left( {{{\bf{W}}^*}} \right) = rank\\
	&\left( {\left( {\lambda _1^*\sum\limits_{k = 1}^K {\frac{{{\bf{H}}_{Ek}^H{\bf{\Theta }}{{\bf{H}}_{Ek}}}}{{\sigma _{Ek}^2}}}  + \lambda _3^*{\bf{H}}_p^H{\bf{\Theta }}{{\bf{H}}_p} + \left( {\eta \zeta  + \lambda _4^*} \right){{\bf{I}}_N}} \right){{\bf{W}}^*}} \right)\\
	&= rank\left( {\left( {\frac{1}{{\sigma _S^2 + tr\left( {{\bf{\Theta }}{{\bf{H}}_S}{{\bf{W}}^*}{\bf{H}}_S^H} \right)}} + \frac{{\lambda _2^*}}{{\sigma _S^2}}} \right){\bf{H}}_S^H{\bf{\Theta }}{{\bf{H}}_S}{{\bf{W}}^*}} \right)\\
	&\le rank\left( {\bf{\Theta }} \right) = 1.
\end{split}
\end{equation}
Since $\eta>0$, ${\bf{W}}^*={\bf{0}}$ is unlikely to be the optimal solution. Therefore, we can obtain $rank\left( {{{\bf{W}}^*}} \right)=1$.

\section{Proof of Proposition 4}
According to the iterative process in (28), we have
\begin{equation}
	\begin{split}
		{f_1}&\left( {{{{\bf{ W}}}^{i + 1}},\eta } \right) - {f_2}\left( {{{\bar \varphi }^{i+1}}} \right) - \frac{{{{\bar \varphi }^{i + 2}} - {{\bar \varphi }^{i+1}}}}{{{{\bar \varphi }^{i+1}}\ln 2}}\\
		=& \mathop {\max }\limits_{\left( {{\bf{W}},\varphi } \right) \in {{\cal R}_2}} {f_1}\left( {{\bf{W}},\eta } \right) - {f_2}\left( {{{\bar \varphi }^{i+1}}} \right) - \frac{{\varphi  - {{\bar \varphi }^{i+1}}}}{{{{\bar \varphi }^{i+1}}\ln 2}}\\
		\ge& {f_1}\left( {{{{\bf{ W}}}^i},\eta } \right) - {f_2}\left( {{{\bar \varphi }^{i+1}}} \right),
	\end{split}
\end{equation}
where ${\cal R}_2$ is the set of feasible solutions to problem (28). Moreover, by making use of (27), we can obtain
\begin{equation}
	\begin{split}
		{f_2}&\left( {{{\bar \varphi }^{i + 2}}} \right) \le {f_2}\left( {{{\bar \varphi }^{i+1}}} \right) + \frac{{{{\bar \varphi }^{i + 2}} - {{\bar \varphi }^{i+1}}}}{{{{\bar \varphi }^{i+1}}\ln 2}}.
	\end{split}
\end{equation}
By substituting (48) into (47), one can get the following formula
\begin{equation}
	\begin{split}
		{f_1}&\left( {{{{\bf{ W}}}^{i + 1}},\eta } \right) - {f_2}\left( {{{\bar \varphi }^{i + 2}}} \right)\\
		\ge& {f_1}\left( {{{{\bf{ W}}}^{i + 1}},\eta } \right) - {f_2}\left( {{{\bar \varphi }^{i+1}}} \right) - \frac{{{{\bar \varphi }^{i + 2}} - {{\bar \varphi }^{i+1}}}}{{{{\bar \varphi }^{i+1}}\ln 2}}\\
		\ge& {f_1}\left( {{{{\bf{ W}}}^i},\eta } \right) - {f_2}\left( {{{\bar \varphi }^{i+1}}} \right).
	\end{split}
\end{equation}
Based on (49), we can easily find that the proposed iterative procedure (28) provides better solutions as the number of iterations increases, which helps make the objective function (28a) keep increasing.

On the other hand, considering CBS's transmit power constraint $tr({\bf{W}})\le P_c^{max}$ and making use of Cauchy-Schwarz inequality $tr({\bf{X}}{\bf{Y}})\le tr({\bf{X}})tr({\bf{Y}})$, one can get the upper bound of the objective function (28a) as
\begin{equation}
	\begin{split}
		{f_1}\left( {{\bf{W}},\eta } \right) - {f_2}\left( {\varphi } \right)
		&\le {\log _2}\left( {1 + \frac{{tr\left( {{\bf{\Theta }}{{\bf{H}}_S}{\bf{WH}}_S^H} \right)}}{{\sigma _S^2}}} \right)\\
		&\le {\log _2}\left( {1 + \frac{{P_c^{\max }tr\left( {{\bf{H}}_S^H{\bf{\Theta }}{{\bf{H}}_S}} \right)}}{{\sigma _S^2}}} \right).
	\end{split}
\end{equation}

According to the formulas (49) and (50), we have verified the convergence of the iterative procedure in (28).

%

\appendices

\ifCLASSOPTIONcaptionsoff
\newpage
\fi




\begin{thebibliography}{1}	
	\bibitem{5}
	W. Zhang, C. Wang, X. Ge, and Y. Chen, “Enhanced 5G Cognitive Radio Networks Based on Spectrum Sharing and Spectrum Aggregation," \textit{IEEE Trans. Commun.}, vol. 66, no. 12, pp. 6304--6316, Dec. 2018.
	
	
	
	
	
    \bibitem{re9}
	C. Wang and H. Wang, “On the Secrecy Throughput Maximization for MISO Cognitive Radio Network in Slow Fading Channels," \textit{IEEE Trans. Inform. Forensics Secur.}, vol. 9, no. 11, pp. 1814-1827, Nov. 2014.
	
	
	
	
	
	
		
	\bibitem{a13}			
   H. Xuemin, W. Jing, W. Cheng-Xiang, and S. Jianghong, “Cognitive radio in 5G: a perspective on energy-spectral efficiency
trade-off,” \textit{IEEE Commun. Mag.}, vol. 52, no. 7, pp. 46--53, July 2014.

	\bibitem{a14}			
   H. Al-Hraishawi and G. A. A. Baduge, “Wireless energy harvesting in cognitive massive MIMO systems with underlay spectrum sharing,” \textit{IEEE Wireless Commun. Lett.}, vol. 6, no. 1, pp. 134--137, Feb. 2017.

	\bibitem{a15}			
 D. Hamza, P. Ki-Hong, M. S. Alouini, and S. Aissa, “Throughput maximization for cognitive radio networks using active cooperation and superposition coding,” \textit{IEEE Trans. Wireless Commun.}, vol. 14, no. 6, pp. 3322--3336, June 2015.

	\bibitem{a16}			
 Q. Zhao, S. Geirhofer, L. Tong, and B. M. Sadler, “Opportunistic spectrum access via periodic channel sensing,” \textit{IEEE Trans. Signal Process.}, vol. 56, no. 2, pp. 785--796, Feb. 2008.







	\bibitem{a21}			
 L. B. Le and E. Hossain, “Resource allocation for spectrum underlay in cognitive radio networks,” \textit{IEEE Trans. Wireless Commun.}, vol. 7, no. 12, pp. 5306--5315, Dec. 2008.

	\bibitem{a22}			
 Y. Cao and C. Tellambura, “Cognitive beamforming in underlay two-way relay networks with multiantenna terminals,” \textit{IEEE Trans. Cogn. Commun. Netw.}, vol. 1, no. 3, pp. 294--304, Sept. 2015.

	\bibitem{a23}			
 W. Lee, “Resource allocation for multi-channel underlay cognitive radio network based on deep neural network,” \textit{IEEE Commun. Lett.}, vol. 22, no. 9, pp. 1942--1945, Sept. 2018.



	\bibitem{TT1}	
A. Kaur and K. Kumar, “Imperfect CSI based Intelligent Dynamic Spectrum Management using Cooperative Reinforcement Learning Framework in Cognitive Radio Networks," \textit{IEEE Trans. Mobile Comput.}, doi: 10.1109/TMC.2020.3026415.

	\bibitem{TT2}	
H. Liao, X. Chen, Z. Zhou, N. Liu and B. Ai, “Licensed and Unlicensed Spectrum Management for Cognitive M2M: A Context-Aware Learning Approach," \textit{IEEE Trans. Cogn. Commun. Netw.}, vol. 6, no. 3, pp. 915--925, Sept. 2020.







	\bibitem{a17}			
D. H. Tashman and W. Hamouda, “An Overview and Future Directions on Physical-Layer Security for Cognitive Radio Networks," \textit{IEEE Netw.}, vol. 35, no. 3, pp. 205--211, May/June 2021.




	\bibitem{a18}			
H. Lei, M. Xu, I. S. Ansari, G. Pan, K. A. Qaraqe, and M. Alouini, “On Secure Underlay MIMO Cognitive Radio Networks With Energy Harvesting and Transmit Antenna Selection," \textit{IEEE Trans. Green Commun. Netw.}, vol. 1, no. 2, pp. 192--203, June 2017.

	\bibitem{a19}			
Z. Shu, Y. Qian, and S. Ci, “On physical layer security for cognitive radio networks," \textit{ IEEE Netw.}, vol. 27, no. 3, pp. 28--33, May--June 2013.

	\bibitem{a20}			
Y. Zou, J. Zhu, L. Yang, Y. Liang, and Y. Yao, “Securing physical-layer communications for cognitive radio networks," \textit{IEEE Commun. Mag.}, vol. 53, no. 9, pp. 48--54, Sept. 2015.







	


	
	\bibitem{add}
	A. D.Wyner, “The wire-tap channel,” \textit{Bell Syst. Technol. J.}, vol. 54, no. 8, pp. 1355--1387, 1975.
	
	
	
	
	\bibitem{a27}			
	Q. Li and L. Yang, “Beamforming for Cooperative Secure Transmission in Cognitive Two-Way Relay Networks," \textit{ IEEE Trans. Inform. Forensics Secur.}, vol. 15, pp. 130--143, 2020.
	
	\bibitem{a24}			
	Y. Dong, A. El Shafie, M. J. Hossain, J. Cheng, N. Al-Dhahir, and V. C. M. Leung, “Secure beamforming in full-duplex MISO-SWIPT systems with multiple eavesdroppers,” \textit{IEEE Trans. Wireless Commun.}, vol. 17, no. 10, pp. 6559--6574, Oct. 2018.
	
	
	
	\bibitem{a25}			
	Q. Shi, C. Peng, W. Xu, M. Hong, and Y. Cai, “Energy efficiency optimization for MISO SWIPT systems with zero-forcing beamforming,” \textit{IEEE Trans. Signal Process.}, vol. 64, no. 4, pp. 842--854, Feb. 2016.
	
	\bibitem{a26}			
    A. E.-Shafie, D. Niyato, and N. A.-Dhahir, “Artificial-noise-aided secure MIMO full-duplex relay channels with fixed-power transmissions,” \textit{IEEE Commun. Lett.}, vol. 20, no. 8, pp. 1591--1594, Aug. 2016.
	


\bibitem{1}
Q. Wu and R. Zhang, “Towards smart and reconfigurable environment: Intelligent reflecting surface aided wireless networks,” \textit{IEEE Commun. Mag.}, vol. 58, no. 1, pp. 106--112, Jan. 2020.









\bibitem{a1}
Y. Han, W. Tang, S. Jin, C. Wen, and X. Ma, “Large intelligent surface-assisted wireless communication exploiting statistical CSI,” \textit{IEEE Trans. Veh. Technol.}, vol. 68, no. 8, pp. 8238--8242, Aug. 2019.

\bibitem{a2}
C. Huang, A. Zappone, G. C. Alexandropoulos, M. Debbah, and C. Yuen, “Reconfigurable intelligent surfaces for energy efficiency in wireless communication,” \textit{IEEE Trans. Wireless Commun.}, vol. 18, no. 8, pp. 4157--4170, Aug. 2019.

\bibitem{a3}	
Q. Wu and R. Zhang, “Intelligent reflecting surface enhanced wireless network via joint active and passive beamforming,”
\textit{IEEE Trans. Wireless Commun.}, vol. 18, no. 11, pp. 5394--5409, Nov. 2019.

\bibitem{a4}	
P. Wang, J. Fang, L. Dai, and H. Li, “Joint Transceiver and Large Intelligent Surface Design for Massive MIMO mmWave Systems,"  \textit{IEEE Trans. Wireless Commun.}, vol. 20, no. 2, pp. 1052--1064, Feb. 2021.

\bibitem{a5}	
P. Wang, J. Fang, X. Yuan, Z. Chen, and H. Li, “Intelligent Reflecting Surface-Assisted Millimeter Wave Communications: Joint Active and Passive Precoding Design," \textit{IEEE Trans. Veh. Technol.}, vol. 69, no. 12, pp. 14960--14973, Dec. 2020.

\bibitem{2}
C. Pan \textit{et al.}, “Multicell MIMO communications relying on intelligent reflecting surfaces,” \textit{IEEE Trans. Wireless Commun.}, vol. 19, no. 8, pp. 5218--5233, Aug. 2020.	

\bibitem{3}
H. Guo, Y. Liang, J. Chen, and E. G. Larsson, “Weighted sum rate maximization for reconfigurable intelligent surface aided wireless networks,” \textit{IEEE Trans. Wireless Commun.}, vol. 19, no. 5, pp. 3064--3076, May 2020.






\bibitem{a6}	
M. Cui, G. Zhang, and R. Zhang, “Secure wireless communication via intelligent reflecting surface,” \textit{IEEE Wireless Commun. Lett.}, vol. 8, no. 5, pp. 1410--1414, Oct. 2019.

\bibitem{a7}	
H. Shen, W. Xu, S. Gong, Z. He, and C. Zhao, “Secrecy rate maximization for intelligent reflecting surface assisted multi-antenna communications,” \textit{IEEE Commun. Lett.}, vol. 23, no. 9, pp. 1488--1492, Sept. 2019.

\bibitem{a8}	
X. Yu, D. Xu, Y. Sun, D. W. K. Ng, and R. Schober, “Robust and Secure Wireless Communications via Intelligent Reflecting Surfaces,"  \textit{IEEE J. Sel. Areas Commun.}, vol. 38, no. 11, pp. 2637--2652, Nov. 2020.

\bibitem{a9}	
S. Hong, C. Pan, H. Ren, K. Wang, and A. Nallanathan, “Artificial-Noise-Aided Secure MIMO Wireless Communications via Intelligent Reflecting Surface," \textit{IEEE Trans. Commun.}, vol. 68, no. 12, pp. 7851--7866, Dec. 2020.


\bibitem{re6}	
L. Dong and H. Wang, “Secure MIMO Transmission via Intelligent Reflecting Surface," \textit{IEEE Wireless Commun. Lett.}, vol. 9, no. 6, pp. 787--790, June 2020.

\bibitem{re7}	
L. Dong and H. Wang, “Enhancing Secure MIMO Transmission via Intelligent Reflecting Surface," \textit{IEEE Trans. Wireless Commun.}, vol. 19, no. 11, pp. 7543--7556, Nov. 2020.	




\bibitem{a10}		
S. Hong, C. Pan, H. Ren, K. Wang, K. K. Chai, and A. Nallanathan, “Robust Transmission Design for Intelligent Reflecting Surface Aided Secure Communication Systems with Imperfect Cascaded CSI," \textit{IEEE Trans. Wireless Commun.}, vol. 20, no. 4, pp. 2487--2501, Apr. 2021.


\bibitem{a11}		
B. Feng, Y. Wu and M. Zheng, "Secure Transmission Strategy for Intelligent Reflecting Surface Enhanced Wireless System," \textit{2019 11th International Conference on Wireless Communications and Signal Processing (WCSP)}, Xi'an, China, 2019, pp. 1--6.

\bibitem{4}
Q. Wu and R. Zhang, “Beamforming optimization for wireless network aided by intelligent reflecting surface with discrete phase shifts," \textit{IEEE Trans. Commun.}, vol. 68, no. 3, pp. 1838--1851, Mar. 2020.








\bibitem{a12}			
Q. Wang, F. Zhou, R. Q. Hu and Y. Qian, “Energy Efficient Robust Beamforming and Cooperative Jamming Design for IRS-Assisted MISO Networks," \textit{IEEE Trans. Wireless Commun.}, vol. 20, no. 4, pp. 2592--2607, Apr. 2021.




	
	\bibitem{a28}			
	J. He, K. Yu, Y. Zhou and Y. Shi, “Reconfigurable Intelligent Surface Enhanced Cognitive Radio Networks," \textit{2020 IEEE 92nd Vehicular Technology Conference (VTC2020-Fall)}, Victoria, BC, Canada, 2020, pp. 1--5.
	
	
	
	
	\bibitem{a29}				
	L. Zhang, C. Pan, Y. Wang, H. Ren, K. Wang and A. Nallanathan, “Robust Beamforming Optimization for Intelligent Reflecting Surface Aided Cognitive Radio Networks," \textit{IEEE Global Communications Conference}, Taipei, Taiwan, 2020, pp. 1--6.
	
	
	
	
	
	
	
	\bibitem{a30}				
	L. Zhang, Y. Wang, W. Tao, Z. Jia, T. Song and C. Pan, “Intelligent Reflecting Surface Aided MIMO Cognitive Radio Systems," \textit{IEEE Trans. Veh. Technol.}, vol. 69, no. 10, pp. 11445--11457, Oct. 2020.
	
	\bibitem{a31}					
	X. Guan, Q. Wu and R. Zhang, “Joint Power Control and Passive Beamforming in IRS-Assisted Spectrum Sharing," \textit{IEEE Commun. Lett.}, vol. 24, no. 7, pp. 1553--1557, July 2020.	
	
	\bibitem{a32}					
	J. Yuan, Y. -C. Liang, J. Joung, G. Feng and E. G. Larsson, “Intelligent Reflecting Surface-Assisted Cognitive Radio System," \textit{IEEE Trans. Commun.}, vol. 69, no. 1, pp. 675--687, Jan. 2021.
	

	\bibitem{re8}					
	L. Dong, H. -M. Wang and H. Xiao, “Secure Cognitive Radio Communication via Intelligent Reflecting Surface," \textit{IEEE Trans. Commun.}, vol. 69, no. 7, pp. 4678--4690, July 2021.
	
	
	\bibitem{v1}
	
	M. El-Halabi, T. Liu, and C. N. Georghiades, “Secrecy capacity per unit cost,” \textit{IEEE J. Sel. Areas Commun.}, vol. 31, no. 9, pp. 1909--1920, Sept. 2013.
	
	
	
	
	
	\bibitem{9}
	A. Mukherjee, S. A. A. Fakoorian, J. Huang, and A. L. Swindlehurst, “Principles of physical layer security in multiuser wireless networks: A survey,” \textit{IEEE Commun. Surveys Tuts.}, vol. 16, no. 3, pp. 1550--1573, 3rd Quart., 2014.



	\bibitem{re4}
    Y. Yang, B. Zheng, S. Zhang and R. Zhang, "Intelligent Reflecting Surface Meets OFDM: Protocol Design and Rate Maximization," \textit{IEEE Trans. Commun.}, vol. 68, no. 7, pp. 4522--4535, July 2020.

	\bibitem{re5}
    B. Zheng and R. Zhang, "Intelligent Reflecting Surface-Enhanced OFDM: Channel Estimation and Reflection Optimization," \textit{IEEE Wireless Commun. Lett.}, vol. 9, no. 4, pp. 518--522, Apr. 2020.


	
	\bibitem{re1}	
	T. J. Cui, M. Q. Qi, X. Wan, J. Zhao, and Q. Cheng, “Coding metamaterials, digital metamaterials and programmable metamaterials,” \textit{Light, Sci. Appl.}, vol. 3, p. e218, Oct. 2014.
	
	\bibitem{re2}	
	N. Kaina, M. Dupré, G. Lerosey, and M. Fink, “Shaping complex microwave fields in reverberating media with binary tunable metasurfaces,” \textit{Sci. Rep.}, vol. 4, p. 6693, Oct. 2014.
	
	\bibitem{re3}	
	P. Nayeri, F. Yang, and A. Z. Elsherbeni, \textit{Reflectarray Antennas: Theory, Designs, and Applications}. Hoboken, NJ, USA: Wiley, 2018.
	

	\bibitem{add10}		
	S. Zargari, A. Khalili and R. Zhang, “Energy Efficiency Maximization via Joint Active and Passive Beamforming Design for Multiuser MISO IRS-Aided SWIPT," \textit{IEEE Wireless Commun. Letters}, vol. 10, no. 3, pp. 557--561, Mar. 2021.
	
	
	
	
	
	
	
	
	
	
	\bibitem{10}
	A. Charnes and W. W. Cooper, “Programming with linear fractional functionals,” \textit{Naval Res. Logistics Quart.}, vol. 9, no. 3--4, pp. 181--186, 1962.
	
	\bibitem{11}
	Z.-Q. Luo \textit{et al.}, "Semidefinite relaxation of quadratic optimization problems," \textit{IEEE Signal Process. Mag.}, vol. 27, no. 3, pp. 20--34, May 2010.
	
	\bibitem{b1}
	J. Hiriart-Urruty and C. Lemarechal, \textit{Convex Analysis and Minimization Algorithms I: Fundamentals}. Springer, Berlin, Heidelberg, 1996.
	
	\bibitem{add1}	
	K. Shen and W. Yu, “Fractional programming for communication systems—part I: power control and beamforming,” \textit{IEEE Trans. Signal Processing}, vol. 66, no. 10, pp. 2616--2630, May 2018.
	
	
	
	\bibitem{12}
	H. H. Kha, H. D. Tuan, and H. H. Nguyen, ``Fast global optimal power
	allocation in wireless networks by local D.C. programming,'' \textit{IEEE Trans. Wireless Commun.}, vol. 11, no. 2, pp. 510--515, Feb. 2012.
	
	\bibitem{13}
	K. Nguyen, L. Tran, O. Tervo, Q. Vu, and M. Juntti, “Achieving energy efficiency fairness in multicell MISO downlink,” \textit{IEEE Commun. Lett.}, vol. 19, no. 8, pp. 1426--1429, Aug. 2015.
	
	
	
\end{thebibliography}
%

\end{document}